\documentclass[journal]{IEEEtran}
\usepackage{graphicx}
\usepackage{amsfonts}
\usepackage{mathrsfs}
\usepackage{amssymb,amsmath}
\usepackage{bm}
\usepackage{algorithm}
\usepackage{algorithmic}
\usepackage{theorem}
\usepackage{array,color}
\usepackage{cite}
\usepackage{stfloats}
\usepackage{subfigure}
\usepackage{array}
\usepackage{multirow}
\usepackage{epstopdf}
\usepackage{flushend}
\newtheorem{prop}{Proposition}
\newenvironment{Proof}[1][Proof]{\begin{trivlist}\item[\hskip \labelsep {\bfseries #1}]}{\end{trivlist}}
\newcommand*{\qed}{\hfill\ensuremath{\blacksquare}}
\newcommand{\tb}{\mathbf}

\newcommand{\edit}[1]{{\color{black}#1}}
\newcommand{\editw}[1]{{\color{black}#1}}
\newcommand{\wj}[1]{{\color{black}#1}}
\newcommand{\guo}{\textcolor{black}}
\newcommand{\wn}{\textcolor{black}}
\newcommand{\revise}{\textcolor{black}}
\begin{document}
\title {Iterative Receivers for Downlink MIMO-SCMA: Message Passing and \edit{Distributed} Cooperative Detection }
\author{\IEEEauthorblockN{Weijie Yuan, Nan Wu, Qinghua Guo, Yonghui Li, Chengwen Xing and Jingming Kuang}
\thanks{

{W.\ Yuan, N.\ Wu, C. Xing and J. Kuang are with the School of Information and Electronics, Beijing Institute of Technology, Beijing, China, (e-mail: \{wjyuan, wunan, chengwenxing, jmkuang\}@bit.edu.cn).

Q. Guo is with School of Electrical, Computer and Telecommunications Engineering, The University of Wollongong, NSW, Australia, (e-mail:  qguo@uow.edu.au).

Y. Li is with the Centre of Excellence in Telecommunications, School of Electrical and Information Engineering, University of Sydney, Sydney, NSW 2006, Australia (e-mail: yonghui.li@sydney.edu.au).
 }

{}

{}
}}

\maketitle

\begin{abstract}
\edit{The rapid development of the mobile communications requires ever higher spectral efficiency. The non-orthogonal multiple access (NOMA) has emerged as a promising technology to further increase the access efficiency of wireless networks. Amongst several NOMA schemes, the sparse code multiple access (SCMA) has been shown to be able to achieve better performance. In this paper, we consider a downlink MIMO-SCMA system
over frequency selective fading channels. For optimal detection, the complexity increases exponentially with the product of the number of users, the number of antennas and the channel length. To tackle this challenge, we propose near optimal \editw{low-complexity} iterative receivers based on factor graph. By introducing auxiliary variables, a stretched factor graph is constructed and a hybrid belief propagation (BP) and expectation propagation (EP) receiver, named as `Stretch-BP-EP', is proposed. Considering the convergence problem of BP algorithm on loopy factor graph, we convexify the Bethe free energy and propose a convergence-guaranteed BP-EP receiver, named as `Conv-BP-EP'. We further consider cooperative network and propose two distributed cooperative detection schemes to exploit the diversity gain, namely, belief consensus-based algorithm and Bregman alternative direction method of multipliers (ADMM)-based method. Simulation results verify the superior performance of the proposed Conv-BP-EP receiver compared with other methods. The two proposed distributed cooperative detection schemes can improve the bit error rate performance by exploiting the diversity gain. Moreover, Bregman ADMM method outperforms the belief consensus-based algorithm in noisy inter-user links.
}

\end{abstract}

\section{Introduction}
The multiple access techniques have been widely used in mobile communications, e.g., frequency division multiple access (FDMA) \cite{rappaport1996wireless},  time division multiple access (TDMA) \cite{falconer1995time}, code division multiple access (CDMA) \cite{ojanpera1998wideband} and orthogonal frequency division multiple access (OFDMA) \cite{5771015}. All these multiple access schemes allocate the data of different users to orthogonal resources in order to avoid multiuser interference. However, the rapid development of mobile Internet and traffic growth demands a higher spectral efficiency and ubiquitous connections in the next generation wireless communication systems \cite{andrews2014will}. \revise{Non-orthogonal multiple access (NOMA) has been recognized as a promising candidate to address these challenges due to its capability to further increase the system capacity \cite{dai2015non,wei2016survey}.}

Several NOMA technologies have been proposed in the literature \cite{7582543,saito2013non,wunder20145gnow,7398134,ding2014performance,ding2016application}, which can be \guo{categorized as}: power domain multiplexing and code domain multiplexing.\footnote{\edit{There are other NOMA schemes, e.g., pattern division multiple access.}} In power domain NOMA\footnote{In \edit{several works}, the term NOMA specifically refers to power domain NOMA.}, the signals from different users are multiplexed with different power coefficients. At the receiver side, successive interference cancellation is \edit{used to perform multiuser detection}. Unlike power domain NOMA, the code domain NOMA \edit{employs} sparse spreading sequences instead of dense \guo{ones} to reduce interference. \revise{Some commonly used code domain NOMA schemes include sparse code multiple access (SCMA)\cite{nikopour2013sparse}, interleave division multiple access (IDMA) \cite{xu2017massive} and multi-user shared access (MUSA)\cite{yuan2016multi}. It is shown in \cite{xu2017massive} that a new spatially coupled IDMA scheme has better performance and lower complexity than SCMA system. Nontheless, compared to IDMA, SCMA can further achieve extra shaping gain due to the optimal design of codebook\cite{dai2015non}.} Therefore, we focus on SCMA technology in this paper.

In SCMA, bit mapping and spreading are combined together and different bitstreams are mapped to different sparse codewords directly. All codewords are selected from a predefined SCMA codebook \edit{set, where} the positions of nonzero elements in different codebooks are distinct. In \cite{taherzadeh2014scma,17504356}, systematic approaches for designing SCMA codebooks were proposed. By exploring the \editw{low-density} codewords, \edit{factor graph} \cite{loeliger2007factor} and message passing algorithm (MPA) \cite{kschischang2001factor} can be \edit{employed} at the receiver with practically feasible complexity. \revise{Several multiuser detectors have been proposed for uplink and downlink SCMA scenarios {based on} MPA. In \cite{7448879}, the authors proposed a shuffled message passing algorithm to accelerate the convergence rate. \cite{mu2015fixed} presented a fixed low complexity detector for uplink SCMA system based partial marginalization. In \cite{chen2016sparse}, a Monte Carlo Markov Chain (MCMC) based SCMA decoder was proposed which features low complexity when the codebook size is large. In \cite{7752784}, the authors proposde a low-complexity decoding algorithm based on list sphere decoding (LSD) and they further developed several methods to reduce the size of the search tree.} \edit{To further} improve spectrum efficiency, SCMA can be combined with \guo{the} \edit{multiple-input multiple-output} (MIMO) system\guo{s}. Different from the orthogonal multiple access, the optimal detection for MIMO-SCMA system \edit{suffers from very} high computational complexity \cite{7841967}. In \cite{liu2016gaussian}, Gaussian distribution is utilized to approximate data symbols and two low-complexity MPA-based detectors were proposed for MIMO-NOMA system. \editw{In wideband communications, channel becomes frequency selective and signal suffers from inter-symbol interference (ISI)\cite{som2011low,7516666}.} \editw{In \cite{wo2006graph} and \cite{haselmayr2012factor}, low-complexity receivers based on MPA were proposed for MIMO systems over frequency selective channels}.

It is well known that \guo{the} belief propagation (BP) method gives the exact marginal when the factor graph is loop free\cite{yedidia2005constructing}. However, for MIMO-SCMA system over frequency selective channels, \guo{due to loops of factor graph representations, BP may fail to converge or easily converge to local minima}. This phenomenon can be interpreted from the perspective of variational free energy\cite{friston2007variational}. The \edit{BP} rules can be derived \edit{from} minimizing constrained Bethe free energy. Unfortunately, when the factor graph contains loops, Bethe free energy is no longer convex and the resultant MPA is not guaranteed to converge. To overcome the nonconvexity of Bethe free energy, several approximation methods have been introduced. \edit{Tree-reweighted BP} \editw{was} studied in \cite{wainwright2005new} and \cite{kolmogorov2006convergent}, which is based on a distribution over spanning trees of the original factor graph. 
\wj{However, not every belief propagation variants can be represented on spanning trees\cite{Weiss2012MAP}.} \revise{Motivated by {the fact} that free energy can be approximated by linear combinations of several local entropies, it is possible to choose certain counting numbers to form a convexified Bethe free energy.} Several works have investigated the choices of counting numbers and presented the condition that ensures the convexity of free energy\cite{Weiss2012MAP,meshi2009convexifying}. \editw{Nevertheless, the conditions have to be specified for different applications, and to the best knowledge of the authors,} \guo{so far there have been no investigations on the condition in receiver design}.

\editw{Moreover, for cooperative network in which users are all
	owed to communication with each other, cooperative detection can be performed by users to achieve diversity gain}. Obviously, the basic idea is to share the measurements of all users with each other. Nevertheless, this mechanism is not realistic due to two reasons: 1) transmitting measurements to a possibly distant user is quite power consuming; 2) a multi-hop routing scheme is required to ensure all measurements are collected, which may not be practical. On the contrary, the distributed processing scheme based on cooperation between users in a network is more attractive since it only requires local computations and communications with neighboring users. The distributed cooperative processing method has been widely used in localization and target tracking problems \cite{wymeersch2009cooperative,meyer2016distributed}. For communications, only few papers considered the strategies for base station or receiver cooperation \wj{but} did not apply them to MPA-based receivers in fading channels\cite{zhu2010distributed,ng2008distributed,ding2015cooperative}.

In this paper, \editw{low-complexity receiver design for} MIMO-SCMA system over frequency selective channels is studied. \editw{We introduce \edit{auxiliary variables} and represent the factorization of \edit{joint posterior distribution} by a stretched factor graph}. \revise{Since directly approximating the probability mass function of discrete {symbols} as Gaussian leads to performance loss\cite{liu2016gaussian}, {we employ expectation propagation (EP) \cite{minka2001expectation} to update the messages obtained from channel decoder}.} \edit{Then, using} Gaussian approximation of extrinsic information, all messages on factor graph can be parameterized into a Gaussian form, which reduces the \edit{computational complexity of message updating significantly}. Moreover, \editw{considering the proposed BP-EP receiver may fail to converge in the MIMO-SCMA scenario due to the loopy factor graph, we convexify the Bethe free energy and propose a convergence-guaranteed BP-EP receiver}.

Secondly, \editw{we consider the cooperation between users to improve the detection performance. Based on the observation that global messages on factor graph can be expressed as the product of several local messages calculated at each user, we are able to perform the cooperative detection in a distributed way}. We first propose a belief consensus-based scheme. As all messages are Gaussian distributed, only their \guo{means and variances} are exchanged between users. \editw{Then considering that the inter-user links are noisy in practice and the convergence speed of belief consensus method will become rather slow, we propose an alternative direction method of multipliers (ADMM)-based algorithm \cite{boyd2011distributed} which aims at minimizing the Kullback-Leibler divergence \cite{kullback1951information} between global message and the product of local messages}. Moreover, the Bregman divergence \cite{frigyik2008functional} is \edit{employed} as the penalty term to enable efficient computations.

In summary, the main contributions of this paper are as follows,
 \begin{itemize}
  	\item \edit{By introducing auxiliary variables and stretched factor graph, we propose a low-complexity BP-EP message passing receiver for MIMO-SCMA detection over frequency selective fading channels.}
    \item \edit{Considering the convergence problem of message passing receiver on loopy factor graph, we propose an appropriate solution of counting numbers to convexify the Bethe free energy and derive a convergence-guaranteed BP-EP receiver.}
    \item \edit{To fully exploit the cooperative gain, two distributed implementations, namely, belief consensus-based method and Bregman ADMM-based method are proposed to perform cooperative detection.}
 \end{itemize}
Finally, the proposed algorithms are evaluated via Monte Carlo simulations. The results demonstrate the superior performance of \edit{the} proposed message passing receivers for MIMO-SCMA system and also show the \guo{great potentials} of using cooperative detection.

The remainder of this paper is organized as follows. In Section II, the system model of the considered MIMO-SCMA is introduced. \editw{Section III develops a low-complexity receiver based on stretched factor graph. In Section IV, we propose a convergence-guaranteed receiver by convexifying the Bethe free energy. Two distributed cooperative detection methods are investigated in Section V. Simulation results \guo{are shown and discussed} in Section VI. Finally, conclusions are drawn in Section VII.}

\emph{Notations:} We use a boldface letter to denote a vector. The superscript $T$ {and} $-1$ denote the transpose and {the} inverse operations{, respectively}; $\mathcal{N}$ denotes the neighboring set of a variable or a function; $\mathcal{G}(m_x,v_x)$ denotes a Gaussian distribution of variable $x$ with mean $m_x$ and variance $v_x$; $|\cdot|$ denotes the modulus of a complex number or the cardinality of a set; $\|\cdot\|_2$ denotes the $\ell^2$ norm; $\propto$ represents equality up to a constant normalization factor; $\tb{x}\backslash x$ denotes all variables in $\tb{x}$ except $x$.

\section{Problem Formulation}
\subsection{System Model}
\begin{figure}[]
\centering
\includegraphics[width=.5\textwidth]{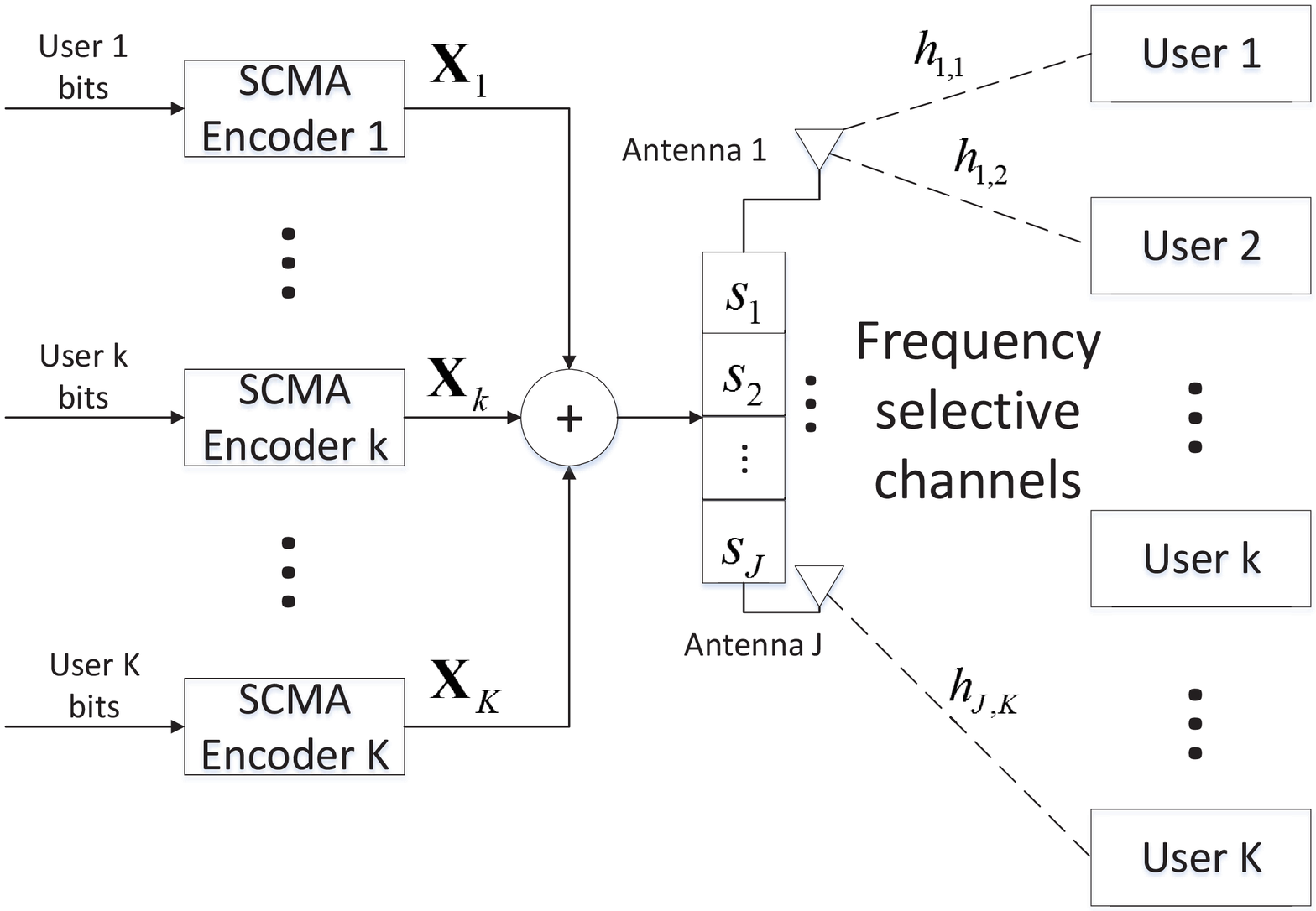}
\caption{\edit{System model for downlink MIMO-SCMA}.}\label{model}
\centering
\end{figure}

We consider a \guo{$K$-user} downlink MIMO-SCMA system, \guo{where each} user is equipped with \guo{a} single antenna and the base station is equipped with $J$ transmit antennas. In the orthogonal multiple access scenario, we usually set $K\leq J$ to avoid inter-user interferences. In the non-orthogonal scenario, $K$ can be greater than $J$ and we define $\varrho=\frac{K}{J}$ as overloading factor. The SCMA encoder is a mapping function \editw{that maps} every $\log_2 M$ coded bits to an $J$-dimensional SCMA codeword. The codewords are selected from a user-specific SCMA codebook of size $M$. Let $\tb{x}_k^n=[x_{k,1}^n,...,x_{k,J}^n]^T$ be the transmitted codeword of user $k$ at time \guo{instant} $n$, which is a sparse vector with $D<J$ nonzero entries. Then the codewords can be multiplexed over $J$ transmit antennas at the base station. The block diagram of the considered system is illustrated in Fig.~ \ref{model}.

Let's denote the transmitted symbol at the $j$th antenna and time instant $n$ by $s_{j}^n$, then $s_j^n$ is given by
\begin{align}
	s_j^n= \sum_{k=1}^K x_{k,j}^n.
\end{align}
In order to capture the sparse feature of SCMA, a binary indicator vector $\tb{f}_k$ is introduced for user $k$, where the $j$th element in $\tb{f}_k$ is given by
\begin{align}
f_{k,j}=\left\{
\begin{array}{cc}
0& x_{k,j}=0\\
1&  x_{k,j}\neq 0.
\end{array}
\right.
\end{align}
The indicate matrix is given as $\tb{F}=[\tb{f}_1,...,\tb{f}_K]$, \guo{where} the nonzero entries in \guo{the} $j$th row denote the conflicting users over the $j$th antenna while the nonzero entries in \guo{the} $k$th column indicate the resources occupied by user $k$.

The signal from base station transmits over frequency selective fading channels with $L$ taps and is received at different users. \guo{With the assumption of} perfect synchronization between the base station and users, the received signal at user $k$ and time instant $n$ can be further written as
\begin{align}\label{ynk}
y_k^n= \sum_{j=1}^J \sum_{l=0}^{L-1} h^l_{j,k} s_j^{n-l}+\omega_k^n,
\end{align}
where $h^l_{j,k}$ is the $l$th tap coefficient of the multipath channel between the $j$th antenna and $k$th user, \guo{and} $\omega_k^n$ is \edit{additive white Gaussian noise} (AWGN) at time instant $n$ with power spectral density $N_0$.

\subsection{Probabilistic Model}
\revise{We further denote $\tb{X}_k$ and $\tb{y}_k$ as transmitted SCMA codewords and received signal samples of the $k$th user, and $\tb{X}$ as the transmitted symbols of all users.} \editw{Assuming perfect channel state information}, each user can perform the optimal maximum \emph{a posteriori} (MAP) detection based on measurement $\tb{y}_k$, which can be expressed as
\begin{align}\label{map}
	\hat{\tb{X}}_k&=\edit{\mathop{\arg\max}_{\tb{X}_k}} p(\tb{X}_k|\tb{y}_k)\nonumber\\&=\edit{\mathop{\arg\max}_{\tb{X}_k}} \int p(\tb{X}|\tb{y}_k) \textrm{d} \tb{X}\backslash \tb{X}_k.
\end{align}
Following Bayesian rules, $p(\tb{X}|\tb{y}_k)$ reads
\begin{align}
p(\tb{X}|\tb{y}_k)\propto p(\tb{X})p(\tb{y}_k|\tb{X})
\end{align}
where $p(\tb{X})$ is the joint \emph{a priori} distribution and $p(\tb{y}_k|\tb{X})$ is the joint likelihood function. Since all transmitted symbols are assumed independent, we have $p(\tb{X})=\prod_{j,k,n} p(x_{k,j}^n)$, where $p(x_{k,j}^n)$ is calculated based on \edit{the log likelihood ratios (LLRs) of coded bits from the output of channel decoder}.

\editw{The computational complexity of the optimal MAP receiver \edit{in} \eqref{map} increases exponentially with the product of the number of users, the number of antennas and the channel length}. In the following, we develop low-complexity message passing receivers for MIMO-SCMA system.

\section{Low-complexity BP-EP Receiver based on Stretched Factor Graph}\label{stretchfg}
\subsection{Factor Graph Representation}
\begin{figure}[]
\centering
\includegraphics[width=.5\textwidth]{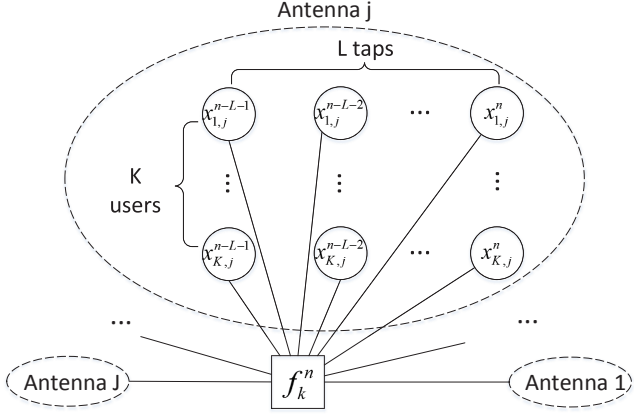}
\caption{\edit{Factor graph representation of the factorization in \eqref{joint}. For ease of exposition, we only plot the variable vertices connected to the factor node $f_k^n$.}}\label{FG}
\centering
\end{figure}

A factor graph is a bipartite graph representing the factorization of a function, which enables efficient computations of marginals. \revise{Since the noise samples at different time \guo{instants} are uncorrelated, the joint likelihood function $p(\tb{y}_k|\tb{X})$ can be factorized as
\begin{align}\label{likelihood}
p(\tb{y}_k|\tb{X})\!&\propto\! \prod_n \! \underbrace{\exp\!\left(\!-\!\frac{\left|y_k^n-\sum_{j=1}^J \sum_{l=0}^{L-1} h^l_{j,k} \sum_{k=1}^K x_{k,j}^{n-l} \right|^2}{2N_0}\!\right)}_{f_n^k}
\end{align}}
Consequently, the joint \emph{a posteriori} distribution can be rewritten as
\begin{align}\label{joint}
	p&(\tb{X}|\tb{y}_k)\propto\prod_n  \prod_{j,k} p(x_{k,j}^n)   \nonumber\\ &\times\!\exp\!\left(\!-\!\frac{\left|y_k^n-\sum_{j=1}^J \sum_{l=0}^{L-1} h^l_{j,k} \sum_{k=1}^K x_{k,j}^{n-l} \right|^2}{2N_0}\!\right)
\end{align}
and can be represented by a factor graph, as depicted in Fig.~\ref{FG}, \revise{where the shorthand notation $f_k^n$ denotes the local likelihood function corresponding to received sample $y_k^n$.} We use squares to denote \edit{factor vertices} and circles to denote \edit{variable vertices}. A variable vertex $x$ is connected to a factor vertex $f$ via an edge if and only if $x$ is a variable \edit{of} the function \editw{$f$}.

\subsection{Stretched Factor Graph and Low-complexity BP-EP Receiver}\label{stdbp}
There are two kinds of messages updated on factor graph, i.e., message from factor vertex $f$ to variable vertex $x$ and message from variable vertex $x$ to factor vertex $f$, denoted by $\mu_{f\to x} (x)$ and $\mu_{x\to \edit{f}} (x)$, respectively. According to the \editw{BP} rules \cite{kschischang2001factor}, $\mu_{f\to x} (x)$ and $\mu_{x\to \edit{f}} (x)$ are given \edit{as}
\begin{align}\label{ftox}
\mu_{f\to x} (x)&\propto\int 	f(\tb{x}) \prod _{x^{'}\in \mathcal{N}(f)\backslash \{x\}}\mu_{x^{'}\to f} (x^{'}) \textrm{d} {x}^{'}\\\label{xtof}
\mu_{x\to f} (x)&\propto\prod_{f^{'}\in \mathcal{N}(x)\backslash\{f\}} \mu_{f^{'}\to x} (x)
\end{align}
and \revise{the belief (`approximate marginal') of variable $x$ is given by}
\begin{align}
b (x)&\propto\prod_{f\in \mathcal{N}(x)} \mu_{f\to x} (x).
\end{align}

\begin{figure}[]
\centering
\includegraphics[width=.5\textwidth]{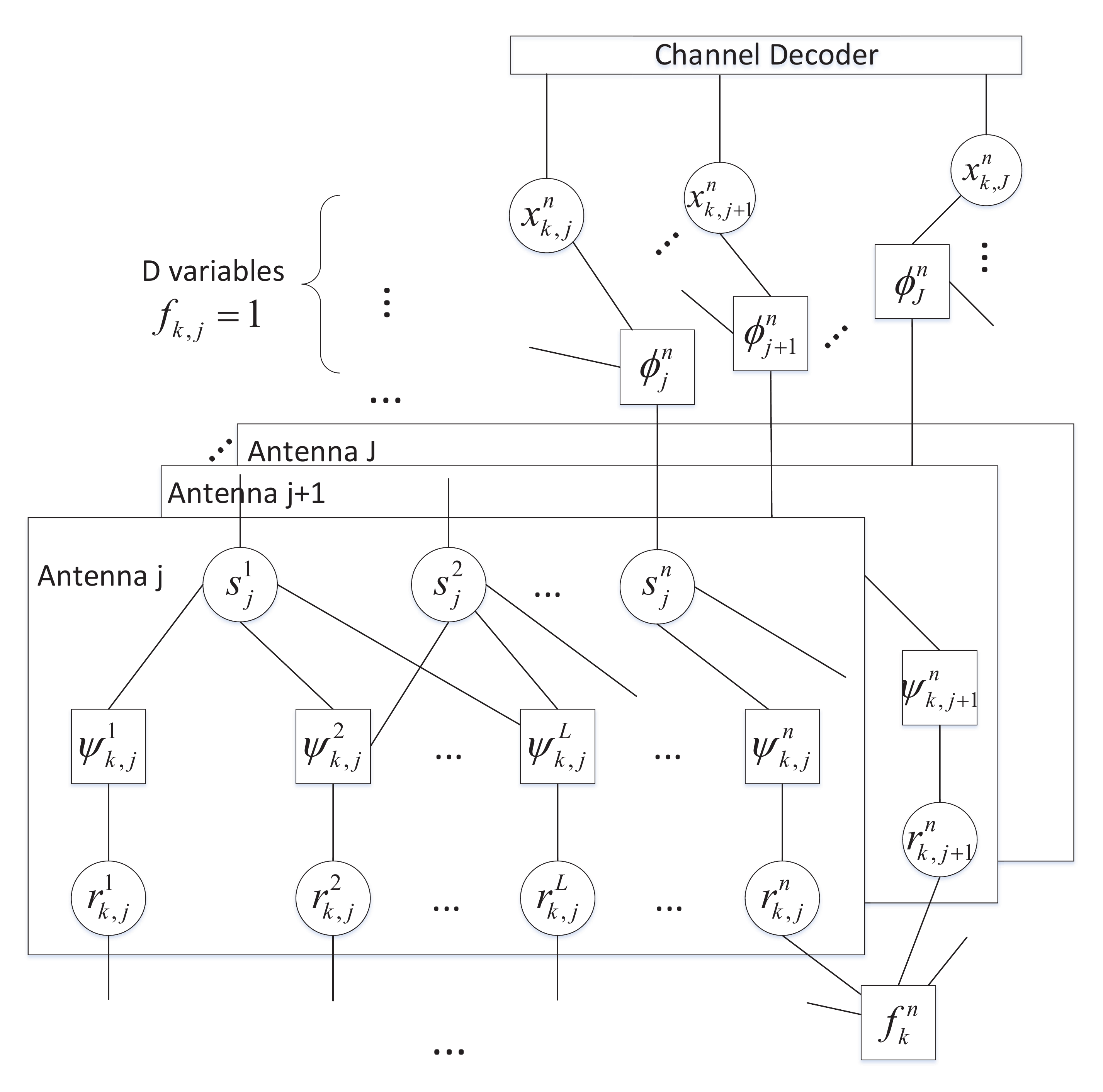}
\caption{Stretched factor graph representation of the considered MIMO-SCMA system.}\label{FG1}
\centering
\end{figure}

Note that in Fig.~\ref{FG}, \edit{$JKL$ variables in total} are connected to a factor node. Following \eqref{ftox}, when calculating the message $\mu_{f_{k}^n\to x_{k,j}^n} (x_{k,j}^n)$, \edit{multi-dimensional integration of $f_{k}^n$ over $(JKL-1)$ variables have to be performed.} \editw{Therefore, the complexity of the BP receiver based on the factor graph in Fig.~\ref{FG} is $\mathcal{O}({N(JKL)^2})$, which is huge in MIMO-SCMA systems}. To tackle this problem, we introduce \edit{auxiliary variables to reduce the number of messages that have to be updated, thereby reducing the complexity of receiver significantly}. \revise{Let $r_{k,j}^n=\sum_{l=0}^{L-1} h_{k,j}^l s_{j}^{n-l}$, and the likelihood function in \eqref{likelihood} can be rewritten as
\begin{align} \label{newlf}
	p(\tb{y}_k|\tb{X})\propto \prod_n \bigg(&\exp\left(-\frac{|y_k^n-\sum_j r_{j,k}^{n}|^2}{2N_0}\right) \psi^n_k  \prod_j \phi_j^n  \bigg)
\end{align}}
where $\psi^n_k=\delta(r_{k,j}^n-\sum_{l=0}^{L-1} h_{k,j}^l s_{j}^{n-l})$ and $\phi_j^n=\delta(s_j^n-\sum_{k=1}^K x_{k,j}^n)$ denote the equality \edit{constraints}. \editw{Based on the \edit{the above} factorization, we are able to stretch multiple variables and construct a novel factor graph as illustrated in Fig.~\ref{FG1}, which is named as `stretched factor graph'}.

Based on the \edit{rules in \eqref{ftox} and \eqref{xtof}, we can update the messages on factor graph as follows}.

\revise{\noindent$\bullet$ Incoming Message $\mu_{x_{k,j}^n\to \phi_j^n} (x_{k,j}^n)$ (EP updating):}

The message $\mu_{x_{k,j}^n\to \phi_j^n} (x_{k,j}^n)$ can be regarded as the prior distribution $p(x_{k,j}^n)$ of symbols, \edit{which is expressed as}
\begin{align}
p(x_{k,j}^n)=\sum_{i=1}^M p_i \delta(x_{k,j}^n-\chi_i)
\end{align}
where $\chi_i$ is the $i$th constellation point, \edit{$p_i$ can be calculated based on the LLRs of bits from the output of channel decoder}. Generally, we can approximate $p(x_{k,j}^n)$ as Gaussian \edit{distribution} by directly matching the first and second order moments. \edit{However}, \edit{this} will lead to performance loss. \edit{To solve this problem,} we resort to EP which matches the moments of belief instead of the prior distribution \cite{7922514}. \revise{Since information from detector is also exploited, a hybrid BP-EP receiver is expected to improve the performance.} Assuming that the message  $\mu_{\phi_j^n\to x_{k,j}^n} (x_{k,j}^n)$ \edit{can be represented by Gaussian distribution} \edit{$\mathcal{G} (m_{\phi_j^n\to x_{k,j}^n}, v_{\phi_j^n\to x_{k,j}^n})$}, the mean and variance of the belief of $x_{k,j}^n$ can be expressed as
\begin{align}
	m_{x_{k,j}^n}=&\frac{1}{2\pi v_{\phi_j^n\to x_{k,j}^n}} \cdot \sum_{i=1}^M \chi_i p_i \exp \left(-\frac{(m_{\phi_j^n\to x_{k,j}^n}-\chi_i)^2}{v_{\phi_j^n\to x_{k,j}^n}}\right),\\ \nonumber 	 v_{x_{k,j}^n}=&\frac{1}{2\pi v_{f\to x_{k,j}^n}}  \cdot \sum_{i=1}^M |\chi_i|^2 p_i \exp \left(-\frac{(m_{\phi_j^n\to x_{k,j}^n}-\chi_i)^2}{v_{\phi_j^n\to x_{k,j}^n}}\right) \\ &-|m_{x_{k,j}^n}|^2.
\end{align}
Consequently, the Gaussian approximation of message $\mu_{x_{k,j}^n\to \phi_j^n} (x_{k,j}^n)$ is given as
\begin{align}\label{epm}
m_{x_{k,j}^n\to \phi_j^n} &=v_{x_{k,j}^n\to \phi_j^n}\left(\frac{m_{x_{k,j}^n}}{v_{x_{k,j}^n}}-\frac{m_{\phi_j^n\to x_{k,j}^n}}{v_{\phi_j^n\to x_{k,j}^n}}\right)\\\label{epv}
v_{x_{k,j}^n\to \phi_j^n}&=\frac{v_{x_{k,j}^n}v_{\phi_j^n\to x_{k,j}^n}}{v_{\phi_j^n\to x_{k,j}^n}-v_{x_{k,j}^n}}.
\end{align}

\revise{\noindent$\bullet$ Messages related to $\phi_j^n$ and $\psi_{k,j}^n$ (BP updating):}

After collecting the messages $\mu_{x_{k,j}^n\to \phi_j^n} (x_{k,j}^n)$, $\forall k$, and assuming that $\mu_{s_{j}^n\to \phi_j^n} (s_{j}^n)\propto \edit{\mathcal{G}}(m_{s_{j}^n\to \phi_j^n},v_{s_{j}^n\to \phi_j^n})$ is obtained, the message $\mu_{\phi_j^n\to x_{k,j}^n} (x_{k,j}^n)$ is updated according to \eqref{ftox}, i.e.,
\begin{align}
\mu_{\phi_j^n\to x_{k,j}^n} (x_{k,j}^n) \propto \int&\delta(s_j^n-\sum_{k=1}^K x_{k,j}^n)\prod_{k^{'}\neq k} \mu_{x_{k^{'},j}^n\to \phi_j^n} (x_{k^{'},j}^n)\nonumber\\&\times\mu_{s_{j}^n\to \phi_j^n} (s_{j}^n)  \textrm{d}s_{j}^n \textrm{d}x_{k^{'},j}^n\nonumber\\
\propto \mathcal{G}&(m_{\phi_j^n\to x_{k,j}^n} ,v_{\phi_j^n\to x_{k,j}^n} )
\end{align}
with the mean and variance given as follows,
\begin{align}\label{mphix}
	m_{\phi_j^n\to x_{k,j}^n}& =m_{s_{j}^n\to \phi_j^n}-\sum_{k^{'}\neq k} m_{x_{k^{'},j}^n\to \phi_j^n}\\\label{vphix}
	v_{\phi_j^n\to x_{k,j}^n} &=v_{s_{j}^n\to \phi_j^n}+\sum_{k^{'}\neq k} v_{x_{k^{'},j}^n\to \phi_j^n}.
\end{align}
The message from $\phi_j^n$ to $s_{j}^n$ can also be expressed as Gaussian distribution with mean and variance \edit{as}
\begin{align}\label{mphis}
	m_{\phi_j^n\to s_{j}^n}&=\sum_{k} m_{x_{k,j}^n\to \phi_j^n}\\\label{vphis}
	v_{\phi_j^n\to s_{j}^n} &=\sum_{k\neq k} v_{x_{k,j}^n\to \phi_j^n}.
\end{align}

Similarly, we can obtain the means and variances of the messages $\mu_{\psi_{k,j}^n\to s_{j}^{n-l}} (x_{k,j}^n)$ and $\mu_{\psi_{k,j}^n\to r_{k,j}^n}$ as
\begin{align}\label{mpsis}
	m_{\psi_{k,j}^n\to s_{j}^{n-l}}& =\frac{1}{h_{k,j}^l}\left(m_{r_{k,j}^n\to \psi_{k,j}^n}-\!\!\!\!\sum_{l^{'}=0,l^{'}\neq l}^{L-1} h_{k,j}^{l^{'}}  m_{s_{j}^{n-l^{'}}\to \psi_{k,j}^n}\right)\\\label{vpsis}
	v_{\psi_{k,j}^n\to s_{j}^{n-l}} &=\frac{v_{r_{k,j}^n\to \psi_{k,j}^n}+\sum_{l^{'}=0,l^{'}\neq l}^{L-1} |h_{k,j}^{l^{'}}|^2  v_{s_{j}^{n-l^{'}}\to \psi_{k,j}^n}}{|h_{k,j}^l|^2}
\end{align}
\vspace{-5mm}
\begin{align}\label{mpsir}
	m_{\psi_{k,j}^n\to r_{k,j}^n}&=\sum_{l=0}^{L-1} h_{k,j}^{l} m_{s_{j}^{n-l}\to \psi_{k,j}^n}\\\label{vpsir}
	v_{\psi_{k,j}^n\to r_{k,j}^n} &=\sum_{l=0}^{L-1} h_{k,j}^{l} v_{s_{j}^{n-l}\to \psi_{k,j}^n}
\end{align}
where $m_{s_{j}^{n}\to \psi_{k,j}^n}$ and $v_{s_{j}^{n}\to \psi_{k,j}^n}$ are given as
\begin{align}\label{mspsi}
	m_{s_{j}^{n}\to \psi_{k,j}^n}&=v_{s_{j}^{n}\to \psi_{k,j}^n}\left(\frac{m_{\phi_j^n\to s_{j}^n}}{v_{\phi_j^n\to s_{j}^n}}+\sum_{l=1}^{L-1} \frac{m_{\psi_{k,j}^{n+l}\to s_{j}^{n}}}{v_{\psi_{k,j}^{n+l}\to s_{j}^{n}}}\right)\\\label{vspsi}
	v_{s_{j}^{n}\to \psi_{k,j}^n}&=\left(v^{-1}_{\phi_j^n\to s_{j}^n}+\sum_{l=1}^{L-1} v^{-1}_{\psi_{k,j}^{n+l}\to s_{j}^{n}} \right)^{-1}.
\end{align}

\revise{\noindent$\bullet$ Messages related to $r_{k, j}^n$ (BP updating):}

Finally, we calculate the message $\mu_{f_k^n\to r_{k, j}^n} (r_{k, j}^n)$. As the message $\mu_{\psi_{k,j}^n\to r_{k,j}^n}$ is known, the outgoing message from $f_k^n$ to $r_{k,j}^n$ can be calculated as
\begin{align}
\mu_{f_k^n\to r_{k, j}^n} (r_{k, j}^n)\propto\int &\exp\left(-\frac{|y_k^n-\sum_j \editw{r_{k,j}^{n}}|^2}{2N_0}\right) \nonumber\\&\times\prod_{j^{'}\neq j} \mu_{r_{k,j^{'}}^n\to f_k^n} (r_{k,j^{'}}^n)\textrm{d}r_{k, j^{'}}^n
\end{align}
which is \edit{in} Gaussian \edit{form} with parameters
\begin{align}\label{mfr}
	m_{f_k^n\to r_{k, j}^n}=y_{k}^n- \sum_{j^{'}\neq j} m_{r_{k,j^{'}}^n\to f_k^n}	\\\label{vfr}
	v_{f_k^n\to r_{k, j}^n}=N_0+  \sum_{j^{'}\neq j} v_{r_{k,j^{'}}^n\to f_k^n}.
\end{align}

\edit{We remark that
by introducing the auxiliary variables, the modified factor graph based on \eqref{newlf} is able to reduce the number of integrated messages to \wj{$(J+K+L-1)$}, which is much lower than $JKL-1$, especially when the number of users and antennas is large. }


\subsection{Algorithm Summary}
 From the expressions in Section III.B, updating the messages relies on other variables. Therefore, the messages on the factor graph are updated iteratively. At the first iteration, since we have no information about the transmitted data symbols, the messages $\mu_{g_{k,j}^n\to x_{k,j}^n} (x_{k,j}^n), \forall k,j,n$ are initialized as zero mean Gaussian distribution. Then in each iteration, the mean and variance values of all messages are updated according to the rules derived as \editw{in} \eqref{epm}-\eqref{vfr}. After determining the message $\mu_{\phi_j^n \to x_{k,j}^n} (x_{k,j}^n)$, the \edit{detector} calculates the extrinsic information \edit{of bits} and feeds them to the channel decoder. After decoding, the channel decoder updates the \edit{LLRs of bits} and starts the next iteration. \revise{The details of \editw{the} proposed stretched factor graph-base BP-EP receiver is summarized in Algorithm~\ref{streBP}.}

 \begin{algorithm}[t]
\allowdisplaybreaks
    \caption{Stretched Factor Graph-based BP-EP Receiver for MIMO-SCMA System over Frequency Selective Channels}
    \begin{algorithmic}[1]
    \STATE {\bf{Initialization}}:
        \STATE The incoming messages are initialized as zero mean Gaussian distribution with \wj{zero mean and} infinite variance;
\FOR{iter=1:$\editw{N_{Iter}}$}
\STATE Compute the means and variances of downward messages according to \eqref{mphis}, \eqref{vphis} and \eqref{mpsir}-\eqref{vspsi};
          \STATE Compute the message from factor vertex $f_k^n$ to variable vertex $r_{k,j}^n$ according to \eqref{mfr}-\eqref{vfr};
       \STATE Compute the means and variances of upward messages according to \eqref{mphix}, \eqref{vphix} and \eqref{mpsis}, \eqref{vpsis};
   \STATE Convert the outgoing messages to LLR and feed them to the channel decoder;
   \STATE \editw{Perform standard BP channel decoding};
   \STATE Calculate the incoming messages using EP as in \eqref{epm} and \eqref{epv};
               \ENDFOR
    \end{algorithmic}\label{streBP}
\end{algorithm}



\section{Convergence-guaranteed BP-EP Receiver} \label{ConBPEP}
\revise{The main drawback of the above algorithm based on standard BP is that it does not guarantee convergence in loopy graphs, e.g., the factor graph illustrated in Fig.~\ref{FG1}. Several papers have investigated this issue, e.g., tree reweighted BP and dampening message method. In this section, we first introduce how to obtain belief propagation message updating rules based on the variational free energy framework \cite{yedidia2005constructing}. Then we propose an iterative message passing receiver with guaranteed convergence by convexifying the Bethe free energy.}
\subsection{Variational Free Energy and Belief Propagation}
\edit{Consider} a joint distribution $p(\tb{x})$ of random variables $\tb{x}=[x_1,...,x_i,...]$ \edit{that} can be factorized into the product of several non-negative functions as
\begin{align}\label{fact}
p(\tb{x})=\prod_a f_a (\tb{x}_a)	
\end{align}
where $a$ is the index of function $f_a$ \guo{with} arguments $\tb{x}_a$ and $\tb{x}_a\triangleq (x_i| i\in \mathcal{N} (a))$. The factorization in \eqref{fact} can be represented by a factor graph. Calculating the marginal distribution $p(x_i)=\sum_{\tb{x}\backslash x_i} p(\tb{x})$ requires the summation over the states of all variables except $x_i$. The variational method is an efficient way to find approximate solutions for the marginals.

Let $b(\tb{x})$ be a positive function approximating $p(\tb{x})$, then the variational free energy is defined as the Kullback-Leibler divergence between $b(\tb{x})$ and $p(\tb{x})$\cite{kullback1951information}, \edit{i.e.,}
\begin{align}\label{vfe}
	F&=\mathbb{D}[{b}(\tb{x})|p(\tb{x})]=\int b(\tb{x})\ln\frac{b(\tb{x})}{p(\tb{x})}\textrm{d}\tb{x}\nonumber\\
      &=	\underbrace{\int b(\tb{x}) \ln  b(\tb{x}) \textrm{d}\tb{x}}_{-H(b)} - 	\int b(\tb{x}) \ln  p(\tb{x}) \textrm{d}\tb{x}
\end{align}
where $H(b)$ is the entropy of $b(\tb{x})$. We aim to find $b(\tb{x})$ which minimizes the variational free energy. To keep \editw{consistency} with the form of $p(\tb{x})$, we employ the Bethe approximation \cite{yedidia2000generalized}, given by
\begin{align}\label{bethe}
b(\tb{x})=\frac{\prod_a b_a(\tb{x}_a)}{\prod_i b_i (x_i)^{|\mathcal{N}(i)|-1}}.
\end{align}
In \eqref{bethe}, $b_a(\tb{x}_a)$ is the joint belief of variables $\tb{x}_a$ and $b_i (x_i)$ is the approximate marginal of $x_i$.

Substituting \eqref{bethe} \edit{into} \eqref{vfe}, the Bethe free energy is obtained as
\begin{align}\label{Fbethe}
F_{B}=&-\sum_a \int 	{b_a(\tb{x}_a)}\ln {f_a(\tb{x}_a)} \textrm{d} \tb{x}_a  +\sum_a \int 	{b_a(\tb{x}_a)}\ln {b_a(\tb{x}_a)} \textrm{d} \tb{x}_a\nonumber\\
&+\sum \limits_{i} (1-|\mathcal{N}(i)|)\int b_i (x_i)\ln b_i (x_i) \textrm{d}x_i .
\end{align}
\textcolor{blue}{Considering the normalization constraint and marginalization constraint, we can construct the corresponding Lagrangian as
\begin{align}\label{lagran}
\mathcal{L}_{B} \triangleq & F_{B}+\sum_i \beta_i \left(\int b_i({x}_i) \textrm{d}{x}_i-1\right)	\nonumber \\&+\sum_a \beta_a \left(\int b_a(\tb{x}_a) \textrm{d}\tb{x}_a-1\right)\nonumber\\
&+\!\sum_i\!\! \sum_{f_a\in \mathcal{N}(i)}\!\!\int\!\! \beta_{ai} (x_i) \left(b_i({x}_i)\!-\!\int \!\!b_a(\tb{x}_a)  \textrm{d} \tb{x}_a\backslash x_i\right) \textrm{d}\tb{x}_i .
\end{align}
}

Setting the partial derivatives of $\mathcal{L}_{B}$ with respect to $\beta_i$, $\beta_a$ and $\beta_{ai}$ to zero result in the normalization constraint and marginalization constraint. Setting $\nabla_{b_a(\tb{x}_a)}=0$ gives
\begin{align} \label{ba}
\ln b_a(\tb{x}_a)= \ln f_a (\tb{x}_a)+ \sum_{i\in \mathcal{N}(a)} \beta_{ai} (x_i) +\beta_a-1.
\end{align}
Similarly, we set $\nabla_{b_i({x}_i)}=0$ and obtain
\begin{align}\label{bi}
({|\mathcal{N}(i)|-1}) \ln b_i({x}_i)=1-\beta_i+ \sum_{f_a\in \mathcal{N}(i)} \beta_{ai} (x_i).
\end{align}

\textcolor{blue}{Taking exponential function on both sides of \eqref{ba} and \eqref{bi} and setting $\exp\left(\beta_{ai} (x_i)\right) =\prod_{f_{a^{'}}\in \mathcal{N}(i)\backslash f_a} \mu_{f_{a^{'}} \to i} (x_i)$, we have } 
\begin{align}\label{baaa}
b_a(\tb{x}_a)&\propto f_a (\tb{x}_a)  \prod_{i\in \mathcal{N}(a)} \prod_{f_{a^{'}}\in \mathcal{N}(i)\backslash f_a} \mu_{f_{a^{'}} \to i} (x_i)  \\\label{biii}
b_i (x_i)&\propto \prod_{f_a\in \mathcal{N}(i)} \mu_{f_{a} \to i} (x_i).
\end{align}
According to the marginalization constraint, integrating all variables in $b_a(\tb{x}_a)$ except $x_i$ gives
\begin{align}\label{bixi}
	b_i(x_i)\propto&\int b_a(\tb{x}_a) \textrm{d} \tb{x}_a\backslash x_i \nonumber \\
\propto &\int f_a (\tb{x}_a)\!\!\prod_{i^{'}\in \mathcal{N}(a)\backslash i} \prod_{f_{a^{'}}\in \mathcal{N}(i^{'})\backslash f_a} \!\!\!\mu_{f_{a^{'}} \to i^{'}} (x_{i^{'}}) \textrm{d} x_{i^{'}}\nonumber\\
&\times\prod_{f_{a^{'}}\in \mathcal{N}(i)\backslash f_a} \mu_{f_{a^{'}} \to i} (x_i).
\end{align}
By comparing \eqref{bixi} with \eqref{biii}, we have
\begin{align}
	&\mu_{f_{a} \to x_i} (x_i) \nonumber \\ \propto &\!\int \!\!f_a (\tb{x}_a)\!\!\!\! \prod_{i^{'}\in \mathcal{N}(a)\backslash i} \prod_{f_{a^{'}}\in \mathcal{N}(i^{'})\backslash f_a} \!\!\!\mu_{f_{a^{'}} \to x_{i^{'}}} (x_{i^{'}}) \textrm{d} \tb{x}_a\backslash x_i
\end{align}
which is the same as the message updating rule in \eqref{ftox}. We further define
\begin{align}
	\mu_{x_i\to f_{a}} (x_i)= \frac{b_i(x_i)}{\mu_{f_{a} \to x_i} (x_i)}\propto  \prod_{f_{a^{'}}\in \mathcal{N}(i)\backslash f_a} \mu_{f_{a^{'}} \to x_i} (x_i)
\end{align}
which is the rule in \eqref{xtof}.
\subsection{Convergence-guaranteed BP-EP Receiver}
In previous subsection, we show that the message updating rules of the standard BP can be derived by minimizing the constrained Bethe free energy. However, in general, the Bethe free energy is  non-convex 
and has several local minima. As a result, BP is not guaranteed to converge.

\revise{Using the definition $H_i(b)= -\int b_i (x_i)\ln b_i (x_i) \textrm{d}x_i$ and $H_a(b)=-\int 	{b_a(\tb{x}_a)}\ln {b_a(\tb{x}_a)} \textrm{d} \tb{x}_a$, the entropy $H(b)$ is rewritten as $H(b)=\sum_i \left(1-\mathcal{N}(i)\right) H_i(b)+ \sum_a H_a(b)$.} As shown by Yedidia \cite{yedidia2005constructing}, the Bethe approximation for entropy can be generalized by linearly combining $H_i(b)$ and $H_a(b)$ as
\begin{align}\label{general}
\tilde{H}(b)=\sum_i c_i H_i(b)+ \sum_a c_a H_a(b)	
\end{align}
where the counting numbers $c_i$ and $c_a$ satisfy $c_i=1-\sum_{f_a \in \mathcal{N}(i)} c_a$. Obviously, $c_a=1$ corresponds to the Bethe approximation. Consequently, the constrained optimization problem in \eqref{lagran} \editw{can be} revised as
\revise{\begin{align} \label{optim1}
&\min\,\,\, 	-\sum_a\!\int \!	{b_a(\tb{x}_a)}\ln {f_a(\tb{x}_a)} \textrm{d} \tb{x}_a \! -\!\sum_a c_{a} H_a(b)\!-\!\sum_i \! c_{i} H_i(b)\nonumber\\
&\texttt{s}.\texttt{t} \, b_i({x}_i)\!=\!\!\!\int \!\!b_a(\tb{x}_a)  \textrm{d} \tb{x}_a\backslash x_i, \!\int \!\!b_a(\tb{x}_a) \textrm{d}\tb{x}_a\!=\!1,
\!\int \!\!b_i({x}_i) \textrm{d}\tb{x}_i\!=\!1
\end{align}}

Based on \eqref{optim1}, we can construct a Lagrangian and derive the corresponding message updating rules. Similarly, we use the notations in \eqref{ftox} and \eqref{xtof} to denote the messages, which are updated as
	 \begin{align}\label{ftoxcov}
{\mu}_{f_a\to x_i} (x_i)&=\left(\tilde{\mu}_{f_a\to x_i} (x_i)\right)^{\gamma_{ai}}\left(\tilde{\mu}_{x_i\to f_a} (x_i)\right)^{\gamma_{ia}-1}\\\label{xtofcov}
{\mu}_{x_i\to f_a} (x_i)&=\left(\tilde{\mu}_{f_a\to x_i} (x_i)\right)^{\gamma_{ai}-1}\left(\tilde{\mu}_{x_i\to f_a} (x_i)\right)^{\gamma_{ia}}
\end{align}
where $\gamma_{ai}={|\mathcal{N}(i)|c_a}/({|\mathcal{N}(i)|c_a+c_i+|\mathcal{N}(i)|-1)}$ and $\gamma_{ia}={|\mathcal{N}(i)|}/{(|\mathcal{N}(i)|c_a+c_i+|\mathcal{N}(i)|-1)}$. The messages $\tilde{\mu}_{f_a\to x_i} (x_i)$ and $\tilde{\mu}_{x_i\to f_a} (x_i)$ are given as
\begin{align}\label{auxftox}
	\tilde{\mu}_{f_a\to x_i} (x_i)&\propto\int  f_a^ \frac{1}{c_a}(\tb{x}_a) \prod _{i^{'}\in \mathcal{N}(a)\backslash i}\mu_{x_i{^{'}}\to f_a} (x_{i^{'}}) \textrm{d} \tb{x}_a\backslash x_i\\\label{auxxtof}
	\tilde{\mu}_{x_i\to f_{a}} (x_i)&\propto \prod_{f_{a^{'}}\in \mathcal{N}(i)\backslash f_a} \mu_{f_{a^{'}} \to i} (x_i).
\end{align}
The detailed derivation of messages \eqref{ftoxcov} and \eqref{xtofcov} is given in \textbf{Appendix}. Note that in \eqref{ftoxcov}-\eqref{auxxtof}, if $c_a=1$ and $c_i=1-|\mathcal{N}(i)|$, the message updating rules \editw{become the standard BP}.

Based on the general form of Bethe free energy, it is possible to find appropriate counting numbers to form a convex free energy. The prominent tree reweighted BP approximates the free energy as the combination of several entropy terms over spanning trees\cite{kolmogorov2006convergent}. Then the edge appearance probability is used as counting number $c_a$ to convexify the free energy. Nevertheless, only \edit{a few} convex free energies can be represented using spanning trees. To this end, we consider a more general condition, as stated in the following proposition.
\begin{prop}
The modified Bethe free energy is convergence-guaranteed if there exist non-negative counting numbers $c_{ia}$, $c_{ii}$ and $c_{aa}$ satisfying
	\begin{align}\label{ca}
	c_a= c_{aa}+\sum_{i\in \mathcal{N}(a)} c_{ia}\\\label{ci}
		c_i= c_{ii}-\sum_{f_a \in \mathcal{N}(i)} c_{ia}
\end{align}
\end{prop}
\wj{
\begin{Proof}
Substituting \eqref{ca} and \eqref{ci} into \eqref{Fbethe} yields
\begin{align}\label{Fconvex}
F_{B}=&-\sum_a \int 	{b_a(\tb{x}_a)}\ln {f_a(\tb{x}_a)} \textrm{d} \tb{x}_a-\sum_a c_{aa} H_a(b)  \nonumber\\
&-\sum_i c_{ii} H_i(b)-\sum_{i,f_a\in \mathcal{N}(i)}c_{ia} (H_a(b)-H_b(b)).
\end{align}
\revise{For the first term on the right hand side of \eqref{Fconvex}, the second order partial derivatives with respect to ${b_a(\tb{x}_a)}$ equals $0$.} Therefore the convexity of $F_B$ is dominated by the convexity of
\begin{align}
F_{conv}=&-\sum_a c_{aa} H_a(b)-\sum_i c_{ii} H_i(b)  \nonumber\\&-\sum_{i,f_a\in \mathcal{N}(i)}c_{ia} (H_a(b)-H_i(b)).	
\end{align}
Since $\frac{\partial^2 H_i(b)}{\partial b_i(x_i)^2}=-\frac{1}{b_i(x_i)}$ and $\frac{\partial^2 H_a(b)}{\partial b_a(\tb{x}_a)^2}=-\frac{1}{b_a(\tb{x}_a)}$ hold, $F_{conv}$ is convex if and only if $c_{aa}\geq 0$, $c_{ii}\geq 0$ and $-\sum_{i,f_a\in \mathcal{N}(i)}c_{ia} (H_a(b)-H_i(b))	$ is convex.

To analyze the convexity of $H_a(b)-H_i(b)$, we borrow the idea from \cite{heskes2006convexity} that $b_i(x_i)=\int \wn{b_a({\tb{x}_a})} d\tb{x}_a\backslash x_i=b_a(x_i)$. Then we can write $F_{ai}=H_i(b)-H_a(b)$ as
\begin{align}
F_{ai}=\int b_a (\tb{x}_a)\ln b_a (\tb{x}_a) \textrm{d}\tb{x}_a-\int b_a(x_i)\ln b_i (x_i) \textrm{d}x_i.
\end{align}
The second order partial derivatives of $F_{ai}$ with respect to $b_a(\tb{x}_a)$ and $b_i(\tb{x}_i)$ are expressed as
\begin{align}\label{F11}
\frac{\partial^2 F_{ai}}{\partial b_a(x_a)^2}&= \frac{1}{b_a(\tb{x}_a)}\\
\frac{\partial^2 F_{ai}}{\partial b_i(x_i)^2}&= -\frac{b_a(x_i)}{(b_i(x_i))^2}\\\label{F22}
\frac{\partial^2 F_{ai}}{\partial b_a (\tb{x}_a)\partial b_i(x_i)}&=\frac{\partial^2 F_{ai}}{\partial b_i(x_i)\partial b_a (\tb{x}_a)}= -\frac{1}{b_i(x_i)}.
\end{align}
\revise{Convexity of $F_{ai}$ is satisfied if the Hessian matrix with components \eqref{F11}-\eqref{F22} is positive semidefinite.} For any beliefs $\tilde{b}=[\tilde{b}_a (\tb{x}_a), \tilde{b}_i(x_i)]$,
\begin{align}
&\int \tilde{b}\left[
\begin{array}{cc}
	\frac{\partial^2 F_{ai}}{\partial b_a(x_a)^2}& \frac{\partial^2 F_{ai}}{\partial b_a (\tb{x}_a)\partial b_i(x_i)}\\
	\frac{\partial^2 F_{ai}}{\partial b_i(x_i)\partial b_a (\tb{x}_a)}& \frac{\partial^2 F_{ai}}{\partial b_i(x_i)^2}
	\end{array}
	\right] \tilde{b}^T \textrm{d} \tb{x}_a\nonumber\\
	=&\int \left(\frac{(\tilde{b}_a (\tb{x}_a))^2}{b_a(\tb{x}_a)}-\frac{2\tilde{b}_a (\tb{x}_a)\tilde{b}_i(x_i)}{\tilde{b}_i(x_i)}+\frac{\tilde{b}_a(x_i)(\tilde{b}_i(x_i))^2}{({b}_i(x_i))^2}\right) \textrm{d} \tb{x}_a\nonumber\\
	=&\int \tilde{b}_a(\tb{x}_a) \left(\frac{(\tilde{b}_a (\tb{x}_a))}{b_a(\tb{x}_a)}-\frac{\tilde{b}_i(x_i)}{{b}_i(x_i)}\right)^2 \textrm{d} \tb{x}_a\geq 0
\end{align}
which indicates that $F_{ai}$ is convex. Therefore, under the conditions \eqref{ca} and \eqref{ci}, the modified Bethe free energy is convergence-guaranteed.\qed
\end{Proof}
}



From \guo{the} above equations, \edit{various} counting numbers can be chosen to obtain different approximate free energy. \edit{Since} we aim at deriving a convergence-guaranteed version of BP algorithm, the modified free energy should be close to the Bethe free energy. Denoting the counting numbers of Bethe approximation as $d_a=1$ and $d_i=1-|\mathcal{N}(i)|$, we need to minimize the $l_2$ norm $\|\tb{c}-\tb{d}\|^2$, which can be formulated as a quadratic program as
\begin{gather}\label{calca}
	\min_{c_{ii},c_{aa},c_{ia}} \sum_a \left(c_{aa}+\sum_{i\in \mathcal{N}(a)} c_{ia}-1\right)^2\\
	\texttt{s}.\texttt{t}\,\,\, c_i=1-\sum_{f_a\in \mathcal{N}(i)}c_a,~\eqref{ca},\eqref{ci}.\nonumber
\end{gather}
The above optimization problem can be easily solved using standard solvers and therefore will not be elaborated here.

It is well known that for real numbers $a$ and $b$, $\left(e^a\right)^b=e^{ab}$. \guo{Therefore}, if $\tilde{\mu}_{f_a\to x_i} (x_i)$ is obtained in Gaussian, $\left(\tilde{\mu}_{f_a\to x_i} (x_i)\right)^{\gamma_{ai}}$ is still Gaussian with the same mean and variance divided by $\gamma_{ai}$. Then similar to Section \ref{stdbp}, Gaussian messages can be derived based on modified message passing rules with $c_i$ and $c_a$. The details of the proposed convergence-guaranteed message passing algorithm is \guo{summarized} in Algorithm \ref{converBP}.
\begin{algorithm}[t]
\allowdisplaybreaks
    \caption{Convergence-guaranteed BP-EP Receiver for MIMO-SCMA System over Frequency Selective Channels}
    \begin{algorithmic}[1]
    \STATE {\bf{Initialization}}:
        \STATE The incoming messages are initialized as Gaussian distribution with \wj{zero mean} and infinite variance;
        \STATE Calculate the counting numbers  by solving \wn{\eqref{calca}};
\FOR{iter=1:$N_{Iter}$}
    \STATE Compute the auxiliary messages from factor vertices to variable vertices according to \eqref{auxftox};
   \STATE Compute the auxiliary messages from variable vertices to factor vertices according to \eqref{auxxtof};
      \STATE Compute the messages $\mu_{f_a\to x_i}$ and $\mu_{x_i\to f_a}$ using \eqref{xtofcov} and \eqref{ftoxcov};
   \STATE Convert the outgoing messages to LLR and feed them to the channel decoder;
   \STATE \wn{Perform standard BP channel decoding};
   \STATE Calculate the incoming messages using expectation propagation as in \eqref{epm} and \eqref{epv};
               \ENDFOR
    \end{algorithmic}
    \label{converBP}
\end{algorithm}

\subsection{Complexity}
The complexities of proposed BP-EP receiver based on original and stretched factor graphs have already been analyzed in Section III. B. \revise{The complexity of proposed convergence-guaranteed BP-EP receiver is also dominated by the number of messages to be integrated.} As the number of integrated messages does not change, the complexity required for message passing is still $\mathcal{O} (N(J+K+L-1)^2)$. Note that in \wn{\eqref{calca}}, a quadratic programming should be solved to determine the counting number, which improves the number of operations. Nevertheless, since \wn{\eqref{calca}} can be done off-line before perform detection, the complexity of proposed convergence-guaranteed BP-EP receiver is $\mathcal{O} (N(J+K+L-1)^2)$. In Table I, the computational complexities of different receivers are compared.

\wj{
\renewcommand\arraystretch{1.3}
\begin{table}[h]
\caption{Computational Complexities of Different Receivers}
\centering
\label{overhead}
\begin{tabular}{|c|c|}
\hline
Receivers & Computational Complexity\\
\hline
MAP & $\mathcal{O} (N\cdot 2^{JKL-1})$\\
\hline
MMSE & $\mathcal{O} ((NJK)^3)$ \\
\hline
\editw{BP-EP (Original Factor Graph)} & $\mathcal{O} (N(JKL-1)^2)$ \\
\hline
\editw{BP-EP (Stretched Factor Graph)} & $\mathcal{O} (N(J+K+L-1)^2)$\\
\hline
\editw{Convergence-guaranteed BP-EP} & $\mathcal{O} (N(J+K+L-1)^2)$\\
\hline
\end{tabular}
\end{table}
}

\section{\edit{Distributed} Cooperative Detection}


\begin{figure}[h]
\centering
\includegraphics[width=.5\textwidth]{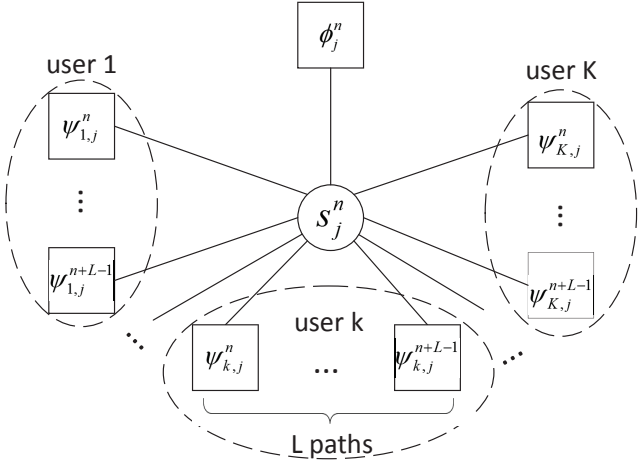}
\caption{Factor graph representation for cooperative detection.}\label{FG2}
\centering
\end{figure}

\editw{We further consider the MIMO-SCMA system in cooperative network where users are allowed to communicate with each other. Since the transmitted symbols $s_j^n$ from base station are received by all the users, cooperative detection can be performed to exploit the diversity gain.}

Thanks to the factor graph framework, it is possible to represent the relationship of $s_{j}^n$ and measurements observed at different users graphically, as illustrated in Fig.~\ref{FG2}. \editw{Note that for each user, the stretched factor graph proposed in Section~\ref{stretchfg} is adopted and the convergence-guaranteed BP-EP algorithm proposed in Section~\ref{ConBPEP} is employed}. Denoting $\mu_{k}(s_j^n)$ as the message to variable vertex $s_j^n$ based on the measurement of user $k$, \edit{we have}
\begin{align}
	\mu_{k}(s_j^n)  =\prod_{l=0}^{L-1} \mu_{\psi_{k}^{n+l} \to s_j^n}(s_j^n).
\end{align}
Having the messages $\mu_{k}(s_j^n),~\forall k$, we can calculate the message $\mu_{s_j^n\to \phi_{j}^n}(s_j^n)$ and then derive the extrinsic information corresponding to data symbol $x_{k,j}^n$. Obviously,
if the measurements are collected by a central unit, it is easy to obtain $\mu_{s_j^n\to \phi_{j}^n}(s_j^n)$ as the product of $\mu_{k}(s_j^n),~\forall k$. However, transmitting measurements to a possibly distant \edit{central unit} \edit{leads to huge power consumption}. On the contrary, using distributed method only requires local communications with neighboring users. By exchanging packets between neighboring users, all users can fully exploit the information related to $s_{j}^n$. In what follows, \edit{two distributed} cooperative detection methods for the considered MIMO-SCMA system are devised.

 \subsection{Belief Consensus-Based Method}
Let $\mathcal{S}_k$ denotes the neighboring set of user $k$, i.e., every user in $\mathcal{S}_k$ can communicate with user $k$. The goal is to determine $\mu_{s_j^n\to \phi_{j}^n}(s_j^n)$ (`global message') at each user with only local processing and communications. The belief consensus method is efficient to compute the product of several local functions over the same variable distributively.

With its local message $\mu_{k}(s_j^n)$ based on measurement $\tb{y}_k$, user $k$ updates its local belief $\rho^{p+1}_{k}(s_j^n)$ according to standard belief consensus recursion, i.e.,
\begin{align}
\rho^{p+1}_{k}(s_j^n)=\rho^{p}_{k}(s_j^n)\prod_{i\in \mathcal{S}_k }\left(\frac{\rho^{p}_{i}(s_j^n)}{\rho^{p}_{k}(s_j^n)}\right)^{\eta}	
\end{align}
where the superscript $p$ denotes the index of consensus iterations and $\eta$ is the update rate. At the first iteration, the local belief is initialized as $\rho^{0}_{k}(s_j^n)=\mu_{k}(s_j^n)$. \edit{In the standard belief consensus, all users share the a constant \guo{updating} rate $\eta$,} which may cause performance loss. The metropolis weight \cite{xiao2006distributed} can be used to solve this problem as
\begin{align}\label{metro}
	\rho^{p+1}_{k}(s_j^n)= \rho^{p}_{k}(s_j^n)^{\eta_{kk}}\prod_{i\in \mathcal{S}_k }{\rho^{p}_{i}(s_j^n)}^{\eta_{ki}}
\end{align}
with the update rate
\begin{align}
\eta_{ik}=\eta_{ki}=\left\{
\begin{array}{cc}
1/\max (|\mathcal{S}_k|,|\mathcal{S}_i|), ~& \textrm{for}\,\, i\neq k\\
1-\sum_{i^{'}\in \mathcal{S}_k} \eta_{i^{'} k},~& \textrm{for}\,\, i= k.
\end{array}	
\right.
\end{align}

With \guo{the assumption of \edit{Gaussian} messages}, \edit{users are able to exchange parameters of the messages instead of the distribution}. In this case, \eqref{metro} can be rewritten as
\begin{align}\label{average}
\bm{\theta}_k^{p+1} &=\eta_{kk} \bm{\theta}_k^{p}+\sum_{i\in \mathcal{S}_k } \eta_{ki} \bm{\theta}_i^{p}
\end{align}
where $\bm{\theta}_k^{p}=[m_{k\to s_j^n}^{{p}}/v_{k\to s_j^n}^{p},1/v_{k\to s_j^n}^{p}]^T$ represents the parameters to be exchanged. Consequently, users only broadcast \wn{$\bm{\theta}_k^p$} to neighboring users.\footnote{It may happen that in a consensus iteration, the link between two users, e.g., user $i$ and $k$ fails. In this case, we use an additional variable $\bar{\bm{\theta}}_{ik}$ to store the parameter received in previous iteration. Then we can use $\bar{\bm{\theta}}_{ik}$ to continue consensus updating.} \revise{For a connected graph that each user has at least one neighbor, after running several consensus iterations, all users reach consensus on the global message, i.e., $\rho^{N_p}_{k}(s_j^n)=\mu_{s_j^n\to \phi_{j}^n}(s_j^n)^{1/K}$, $\forall k$ \cite{meyer2016distributed}.} However, when exchanging packets between users, inter-user links may suffer from additive noise, which causes the variance of $\bm{\theta}_k$ growing unbounded. To tackle this problem, a vanishing parameter is introduced and \eqref{average} is replaced by
\begin{align}\label{modify}
\bm{\theta}_k^{p+1} &= \bm{\theta}_k^{p}+\alpha^{p} \sum_{i\in \mathcal{S}_k } \eta_{ki} \left(\bm{\theta}_i^{p}+\bm{\omega}_{ki}^{p}-\bm{\theta}_k^{p}\right)	
\end{align}
where $\bm{\omega}_{ki}$ is the additive noise on link $i\to k$ and $\alpha^p$ is the vanishing parameter. However, \edit{since $\alpha^p$ decreases monotonically as the increase of $p$}, the convergence speed of \eqref{modify} will be rather slow.

\subsection{Bregeman ADMM-Based Method}
To reach the consensus, the product of all local beliefs should be as \edit{close} as possible to the global message. Motivated by this, the distributed processing problem can be reformulated as \guo{the} minimization of the Kullback-Leibler divergence with the constraint $\rho_{k}(s_j^n)=\rho_{i}(s_j^n)$:
\begin{align}
&\min_{\bm{\rho}}~~\mathbb{D}[\mu_{s_j^n\to \phi_{j}^n}(s_j^n)|\prod_k \rho_{k}(s_j^n)]\\
&\texttt{s}.\texttt{t}~~\rho_{k}(s_j^n)=\rho_{i}(s_j^n), ~\forall~k,i\in \mathcal{S}_k. \nonumber
\end{align}
Considering \guo{that} $\rho_{k}(s_j^n)$ can be characterized by $\bm{\theta}_k$, the objective functional can be minimized subject to $\bm{\theta}_k=\bm{\theta}_i$. For decoupling purpose, we define a set of additional variable $\bm{\pi}_{ki}$ for each inter-user link, through which the optimization problem can be rewritten as 
\begin{align}\label{admm}
&\min_{\bm{\theta}}~~\mathbb{D}[\mu_{s_j^n\to \phi_{j}^n}(s_j^n)|\prod_k \rho_{k}(s_j^n)]\\
&\texttt{s}.\texttt{t}~~\bm{\theta}_{k}=\bm{\pi}_{k}, ~\bm{\theta}_{i}=\bm{\pi}_{k}, ~\forall k,i\in \mathcal{S}_k.\nonumber
\end{align}

\edit{The optimization problem in \eqref{admm} subject to equality constraints can be solved by ADMM.}
In each iteration, ADMM updates variables in a block coordinate fashion by solving the augmented Lagrangian of \eqref{admm}, which is defined as
\begin{align}\label{lagadmm}
\mathcal{L}(\bm{\theta},\bm{\pi},\bm{\lambda})&=	\mathbb{D}[\mu_{s_j^n\to \phi_{j}^n}(s_j^n)|\prod_k \rho_{k}(s_j^n)] \nonumber \\&+\sum_k \sum_{i\in \mathcal{S}_k}\left(\bm{\lambda}_{kk}^T (\bm{\theta}_{k}-\bm{\pi}_{k})+ \bm{\lambda}_{ki}^T (\bm{\theta}_i-\bm{\pi}_{k})\right)\nonumber\\&+\frac{c}{2}\sum_k \sum_{i\in \mathcal{S}_k}\left(\|\bm{\theta}_k-\bm{\pi}_{k}\|_2^2+\|\bm{\theta}_i-\bm{\pi}_{k}\|_2^2\right)
\end{align}
where $c>0$ denotes the penalty coefficient and $\bm{\lambda}$ is the associated Lagrangian multipliers. The quadratic penalty term may result in a high complexity when solving \eqref{lagadmm}. To this end, we resort to replacing the quadratic penalty term by Bregman divergence to generalize ADMM. Let $\varepsilon$ be a continuously differentiable and strictly convex function, namely, Bregman function. The Bregman divergence is defined as
\begin{align}\label{bregman}
B_{\varepsilon} ({x},{y})	=\varepsilon(x)-\varepsilon(y)-\langle x-y, \nabla_{\varepsilon} (y) \rangle
\end{align}
where $\nabla_{\varepsilon} (y)$ is the gradient of $\varepsilon$ and $\langle\cdot \rangle$ denotes the inner product.

Based on Bregeman divergence, we have the following augmented Lagrangian
\wj{
\begin{align}\label{lagbadmm}
\mathcal{L}_{Breg}(\bm{\theta},\bm{\pi},\bm{\lambda})=&	\mathbb{D}[\mu_{s_j^n\to \phi_{j}^n}(s_j^n)|\prod_k \rho_{k}(s_j^n)]\nonumber \\&+\sum_k \sum_{i\in \mathcal{S}_k}\left(\bm{\lambda}_{ki}^T (\bm{\theta}_k-\bm{\pi}_{k})+ \bm{\lambda}_{ik}^T (\bm{\pi}_{k}-\bm{\theta}_i)\right)\nonumber\\&+{c}\sum_k \sum_{i\in \mathcal{S}_k}B_{\varepsilon} (\bm{\theta}_i, \bm{\pi}_{k}).
\end{align}
}
Following the Bregman ADMM, we minimize $\mathcal{L}_B$ with respect to one set of variables given the others. At the \edit{$(p+1)$th} iteration, the updates for Bregman ADMM can be described as
\begin{align}\label{update1}
\bm{\theta}^{p+1}&=\arg \min_{\bm{\theta}} \mathcal{L}_{Breg}(\bm{\theta},\bm{\pi}^{p},\bm{\lambda}^{p})	\\
\bm{\pi}^{p+1}&=\arg \min_{\bm{\pi}} \mathcal{L}_{Breg}(\bm{\theta}^{p+1},\bm{\pi},\bm{\lambda}^{p})\\\label{update3}
\bm{\lambda}^{p+1}_{ki}&=\bm{\lambda}^{p}_{ki}+c (\bm{\theta}^{p+1}_{i}-\bm{\pi}^{p+1}_{ki}).
\end{align}

Actually, a series of $\varepsilon$ can be chosen to form different Bregman divergences. For efficient computations, we need to find an appropriate Bregman function. In our case, since the messages \edit{are in Gaussian form}, we employ log partition function as \edit{Bregman function}. Consequently, the Bregman divergence between two variables is equivalent to the Kullback-Leibler divergence between two Gaussian distributions characterized by these two variables, i.e., $B_{\varepsilon} (\tb{a},\tb{b})=\mathbb{D}[f(\tb{x}|\tb{a})|f(\tb{x}|\tb{b})]$. Then \eqref{update1}-\eqref{update3} can be analytically derived as
\begin{align}\label{update4}
\bm{\theta}^{p+1}_k&=\frac{\bm{\theta}_k^0+\sum_{ i\in \mathcal{S}_k\cup k}\left(\bm{\lambda}^{p}_{ki}+c\bm{\pi}^{p}_i\right)}{1+c(|\mathcal{S}_k|+1)}	\\
\bm{\pi}^{p+1}_k&=\frac{\sum_{ i\in \mathcal{S}_k\cup k} \left(c\bm{\theta}^{p+1}_i-\bm{\lambda}^{p}_{ki} \right)}{c(|\mathcal{S}_k|+1)}\\
\bm{\lambda}^{p+1}_{ki}&=\bm{\lambda}^{p}_{ki}+{c} (\bm{\theta}^{p+1}_{k}-\bm{\pi}^{p+1}_{i})\label{update5}.
\end{align}
By exchanging $\bm{\theta}^{p}_k$ and $\bm{\pi}^{p}_{k}$ in the network, all users can obtain the message $\mu_{s_j^n\to \phi_{j}^n}(s_j^n)$ in a distributed fashion. \revise{As the penalty parameter $c$ will affect the convergence speed, in this paper we consider that $c$ are different for different users which is varying in each iteration\cite{song2015fast}, given as
\begin{align}
c_k^{p+1}=\left\{
\begin{array}{cc}
c_k^p\cdot (1+\tau)& ,\textrm{if} \|\bm{\epsilon}_k^t\|_2>\kappa \|\bm{\iota}_k^t\|_2\\
c_k^p\cdot (1+\tau)^{-1} &,\textrm{if} \|\bm{\iota}_k^t\|_2>\kappa \|\bm{\epsilon}_k^t\|_2\\
c_k^p& ,\textrm{otherwise}
\end{array}
\right.
\end{align}
where $\|\bm{\epsilon}_k^t\|_2$ and $\|\bm{\iota}_k^t\|_2$ are the primal and dual residuals, defined as
{
$\|\bm{\epsilon}_k^t\|_2=\|\bm{\theta}_k^p-\bar{\bm{\theta}}_k^p\|_2$, $\|\bm{\iota}_k^t\|_2=\|\bar{\bm{\theta}_k}^p-\bar{\bm{\theta}}_i^{p-1}\|_2$, $\bar{\bm{\theta}_k}^p=\frac{1}{|\mathcal{S}_k|}\sum_{i\in{\mathcal{S}_k}} \bm{\theta}_i^p$. Typical values of $\kappa$ and $\tau$ are suggested as constant $\kappa=10$ and $\tau=1$.}}

\begin{prop}
Following Bregman ADMM updates rules, all local parameters can reach consensus on the global parameter.   	
\end{prop}
\wj{
\begin{Proof}
Note that for all $k$, the second order partial derivative of the functional $\mathbb{D}[\mu_{s_j^n\to \phi_{j}^n}(s_j^n)|\prod_k \rho_{k}(s_j^n)]$ satisfies
$$
\frac{\partial^2 \mathbb{D}[\mu_{s_j^n\to \phi_{j}^n}(s_j^n)|\prod_k \rho_{k}(s_j^n)]}{\partial  \rho_{k}(s_j^n)^2}=\frac{1}{ \rho_{k}(s_j^n)}>0
$$
which shows the optimization objective is convex. Moreover, since $\varepsilon$ is strictly convex, the Bregman penalty term is also convex. This implies that the optimization problem we are solving is convex and the convergence is guaranteed. \qed	
\end{Proof}

After several iterations, the local belief \edit{$\rho^{N_p}_{k}(s_j^n)$, $\forall k$} characterized by $\bm{\theta}^{N_c}_k$ is guaranteed to converge to the global message $\mu_{s_j^n\to \phi_{j}^n}(s_j^n)$.} \revise{Compared to belief consensus-based method, it can be seen the Bregman ADMM-based algorithm requires to transmit an additional variable, which doubles the communication overhead.}

If the inter-user links suffer from the additive noise, the updates \eqref{update4} and \eqref{update5} can be interpreted as stochastic gradient updates, whose variances have been proved to be bounded values \cite{bertsekas1989parallel}.


\subsection{Algorithm Summary}
For the distributed cooperative detection, \guo{we assume} that each user has obtained the message $\mu_{k\to s_j^n} (s_j^n), \forall k,j,n$ based on its local measurements, \guo{and} the goal is to obtain the product of all users' messages distributively. To start with the distributed algorithm, the local belief $\rho^0_{k} (s_j^n)$ is initialized as $\mu_{k\to s_j^n} (s_j^n)$. Then according to belief consensus-based method and Bregman ADMM-based method, all users update its local belief to reach agreement on the global message. \revise{Also, with the advantage of Gaussian distribution, only few parameters are exchanged in cooperative detection. For both schemes, the complexity is $\mathcal{O}(N)$, which linearly increases with the number of users, {making} them attractive in practical applications.} The proposed \edit{distributed} cooperative detection methods are summarized in \edit{Algorithm 3}.

  \textcolor{black}{\begin{algorithm}[t]
\label{DEM}
\allowdisplaybreaks
    \caption{Belief Consensus and Bregman ADMM-based Methods for Distributed Cooperative Detection}
    \begin{algorithmic}[1]
        \STATE Each user computes message $\mu_{k\to s_j^n} (s_j^n), \forall k,j,n$ based on its local measurements.
       \STATE\hspace{-3mm} \textbf{Enter} cooperative detection
   \STATE Initialize $\rho_k^0 (s_j^n)$ as $\mu_{k\to s_j^n} (s_j^n)$
     \FOR{p=1:$N_{p}$}
    \STATE \wn{Each user broadcasts the parameters $\bm{\theta}^p_k$ (\emph{Belief Consensus}) or $\bm{\theta}^p_k, \bm{\pi}^p_k, \forall k$ (\emph{Bregman ADMM}) to its neighboring users;}\\
    \STATE \wn{Each user update its local parameters using \eqref{average} (\emph{Belief Consensus}) or \eqref{update4}- \eqref{update5} (\emph{Bregman ADMM}) ;}
             \ENDFOR
    \STATE Calculate the message $\mu_{s_j^n\to \phi_{j}^n}(s_j^n)$ at all receivers;
     \STATE\hspace{-3mm} \textbf{Exit} cooperative detection
\STATE Computes other messages on factor graph with $\mu_{s_j^n\to \phi_{j}^n}(s_j^n)$.
    \end{algorithmic}
\end{algorithm}
}

\section{Simulation Results}
\edit{We evaluate the performance of the proposed receivers by Monte Carlo simulations and compare them with several state-of-the-art methods. Consider a MIMO-SCMA system with $J=4$ antennas, $K=6$ users, $D=2$ nonzero entries in each codeword and $M=4$. Therefore, the overloading factor is $\varrho=150\%$. The SCMA codebook is designed according to \cite{nikopour2013sparse} with the indicator matrix $\tb{F}$ defined as }
\begin{align}
\tb{F}=\left[
\begin{array}{cccccc}
1 & 0 & 1 & 0 & 1 & 0\\
1 & 1 & 0 & 0 & 0 & 1\\
0 & 1 & 1 & 1 & 0 & 0\\
0 & 0 & 0 & 1 & 1 & 1
\end{array}
\right]
\end{align}
A 5/7-rate LDPC code is employed with variable and check node degree distributions being $v(X)=0.0005+0.2852X+0.2857X^2+0.4286X^3$ and $c(X)=0.0017X^9+0.9983X^{10}$. Quadrature phase shifting key (QPSK) is utilized as the mother modulation scheme.
The channel is frequency selective with $L=10$ taps, and each tap is independently generated according to the distribution $h^{l}_{k,j}\sim \mathcal{N}(0,q^l)$, $\forall k, j$. The normalized power delay profile is $q^l=\frac{\exp(-0.1l)}{\sum{q^l}}$. The maximum number of iterations is $N_{Iter}=10$. The simulation results are averaged from 1000 independent Monte Carlo trails.

\begin{figure}[t]
\centering
\includegraphics[width=.46\textwidth]{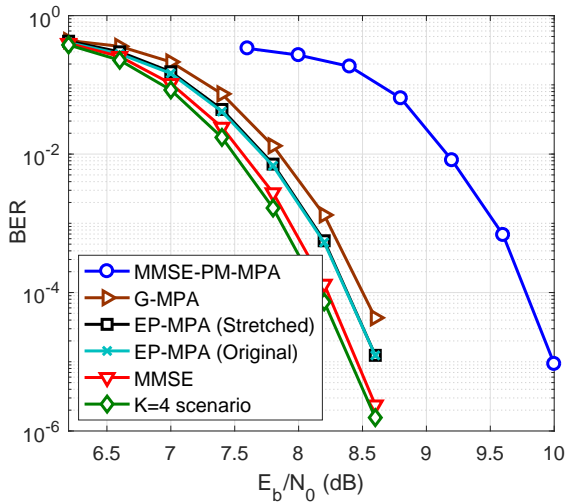}
\revise{\caption{BER performance of different algorithms for MIMO-SCMA system.}}\label{fig1}
\centering
\end{figure}
{\revise{In Fig.~\ref{fig1}, the bit error rate (BER) performance of the proposed stretched factor graph-based BP-EP algorithm (denoted as `Stretch-BP-EP') in Section III.B is plotted.} For comparison, we also \guo{include} the performance of the minimum mean squared error (MMSE)-based method, Gaussian approximated BP (denoted as `GaussAppro-BP') algorithm and a combined MMSE-PM-MPA algorithm\footnote{Gaussian approximated BP is also based on the proposed stretched factor graph. However, the extrinsic information of data symbols are approximated by Gaussian directly, instead of approximating the belief as that in EP. The combined MMSE-PM-MPA receiver first performs the MMSE-based MIMO equalization and then use PM-MPA \cite{mu2015fixed} for SCMA decoding.}. A $K=4$ scenario in which information of different users \guo{is} transmitted using different antennas is also considered as the performance bound (the coding and modulation scheme \guo{are} assumed to be identical to SCMA). It is observed from Fig.~\ref{fig1} that MMSE-PM-MPA method suffers from significant performance loss. This is because that MMSE detector can only output hard information for the PM-MPA-based SCMA detector. \revise{The proposed Stretch-BP-EP algorithm slightly outperforms GaussAppro-BP and performs close to the MMSE-based method. However, the complexity of the proposed algorithm is significantly lower than that of the MMSE-based method.} Moreover, the proposed SCMA system has similar performance compared with the $K=4$ scenario, while the former is able to support 50\% more users.}


\begin{figure}

\centering
\subfigure[Stretched factor graph-based BP-EP algorithm (Stretch-BP-EP)]{\label{fig21}
\includegraphics[width=.46\textwidth]{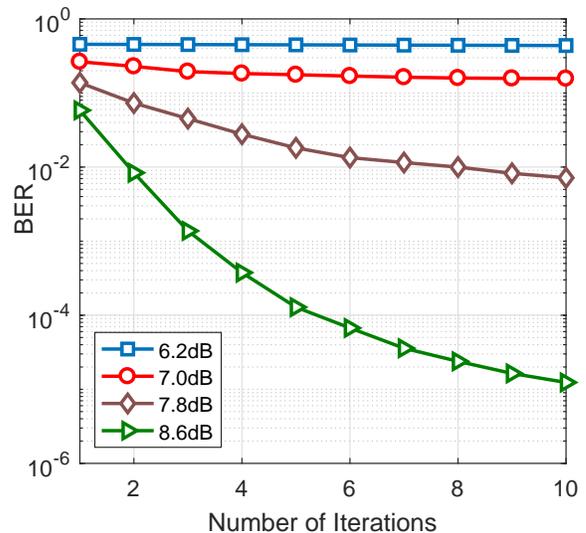}}
\subfigure[Convergence-guaranteed BP-EP algorithm (Conv-BP-EP)]{\label{fig22}
\includegraphics[width=.46\textwidth]{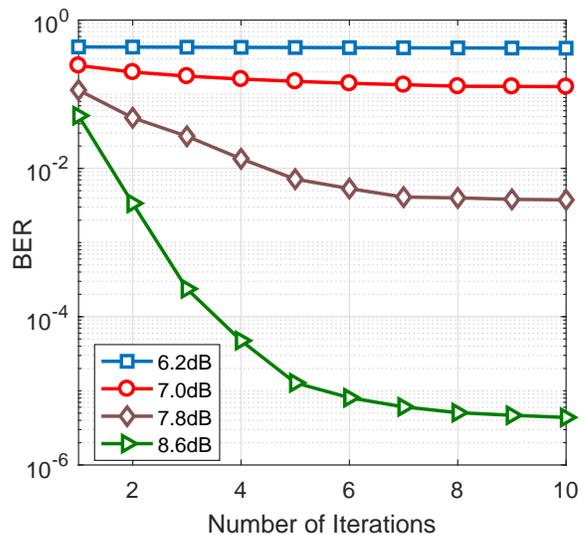}}
\revise{\caption{Impact of the number of iterations on BER performance of Stretch-BP-EP and Conv-BP-EP algorithms.}}\label{fig2}
\centering
\end{figure}

\revise{BER performance of the Stretch-BP-EP method and the proposed convergence-guaranteed BP-EP (denoted as `Conv-BP-EP') algorithm are compared in Fig.~\ref{fig2} at different values of $E_b/N_0$.} It is seen that performance of both algorithms improve as the number of iteration increases. After several iterations, the performance gain of both methods become marginal. By comparing Fig. \ref{fig21} with Fig. \ref{fig22}, we can observe that Conv-BP-EP algorithm converges faster than the Stretch-BP-EP method.
This results can be explained by the fact that Stretch-BP-EP algorithm may converge to the local minima of variational free energy while Conv-BP-EP is guaranteed to converge to the global minimum, which demonstrates the superiority of the proposed Conv-BP-EP method.


In the following, we evaluate the performance of the proposed distributed cooperative detection schemes. \revise{Consider six users uniformly distributed on a $20\times 20 m^2$ unit square. Two users can communicate and exchange information if and only if their distance is less than $d=10m$.} The channels between users are modeled as AWGN and set to be the same for all links. The vanishing parameter for belief consensus is set to a typical value $\alpha^p=\frac{1}{p}$.

\begin{figure}[t]
\centering
\includegraphics[width=.48\textwidth]{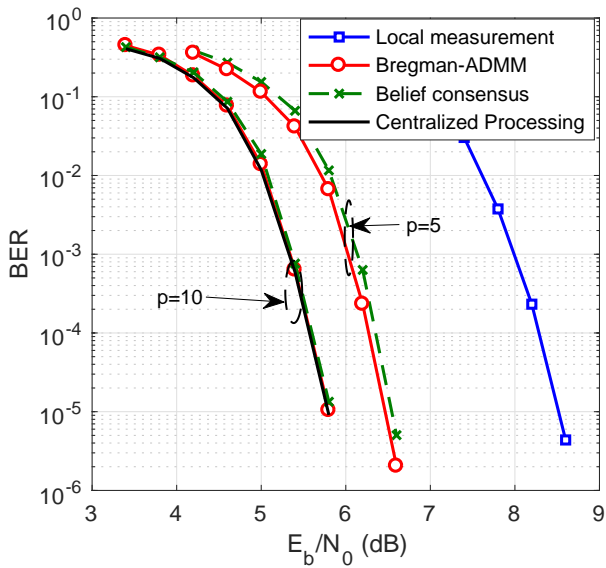}
\caption{BER performance of the proposed distributed cooperative detection schemes with $p=5$ and $p=10$.}\label{fig4}
\centering
\end{figure}

\edit{In Fig.~\ref{fig4}, performance of the proposed two distributed cooperative detection schemes with perfect inter-user links are evaluated. As a benchmark, the BER performance of a centralized scheme is also plotted\footnote{Remark that only the measurements from connected users are collected at a central unit for fair comparison.}. Two values of the number of consensus iterations are considered, i.e. $p=5$ and $p=10$. For comparison, the averaged BER performance of all users based on their local measurements as in Fig. \ref{fig22} is also included. It is observed that, by performing cooperative detection, BER performance can be significantly improved, which reveals that diversity gain can be achieved by exchanging information between neighboring users. By comparing the belief consensus-based method and the Bregman ADMM-based method, we can see that, with perfect inter-user links assumption, both methods \guo{deliver} similar BER performance at $p=5$ and $p=10$. Moreover, after $10$ iterations, both methods attain the performance of centralized processing.}


\begin{figure}[t]
\centering
\includegraphics[width=.505\textwidth]{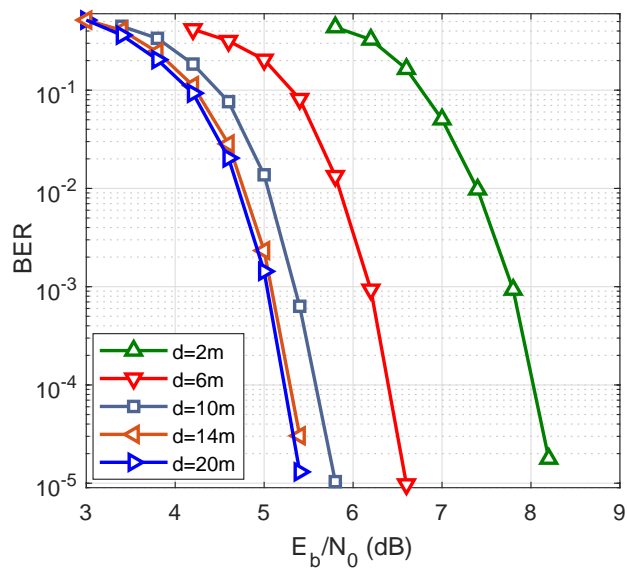}
\caption{Impact of communication range on the BER performance of Bregman ADMM-based method.}\label{fig5}
\centering
\end{figure}

\revise{Since the maximum communication range of inter-user link is critical to the power consumption of users, we compare BER performance of Bregman ADMM-based algorithm with different communication ranges $d=2m$, $d=6m$, $d=10m$, $d=14m$ and $d=20m$. Obviously, increasing $d$ will result in more neighboring users and higher probability of being a fully connected network. As shown in Fig.~\ref{fig5}, BER performance improves as $d$ increases. However, the performance gain becomes smaller when $d$ is large enough. Considering that the power consumption will increase exponentially as $d$ increases, we can trade off between the power cost and BER performance in practice.}

%

\begin{figure}[t]
\centering
\includegraphics[width=.48\textwidth]{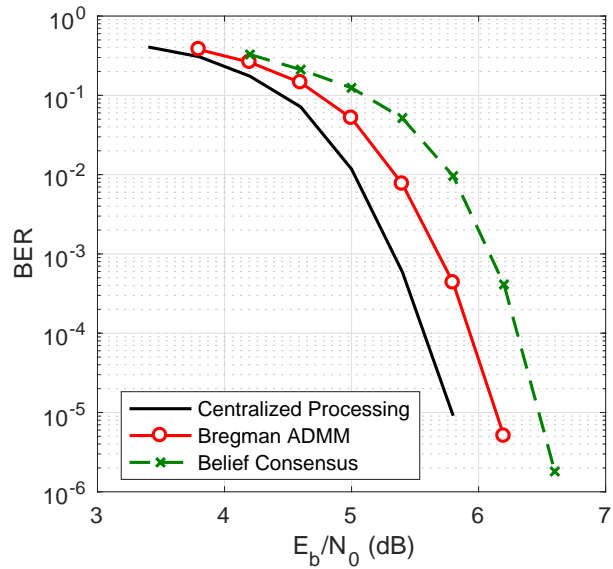}
\caption{BER performance of the proposed distributed cooperative detection schemes with noisy inter-user links.}\label{fig7}
\centering
\end{figure}
\begin{figure}[t]
\centering
\includegraphics[width=.48\textwidth]{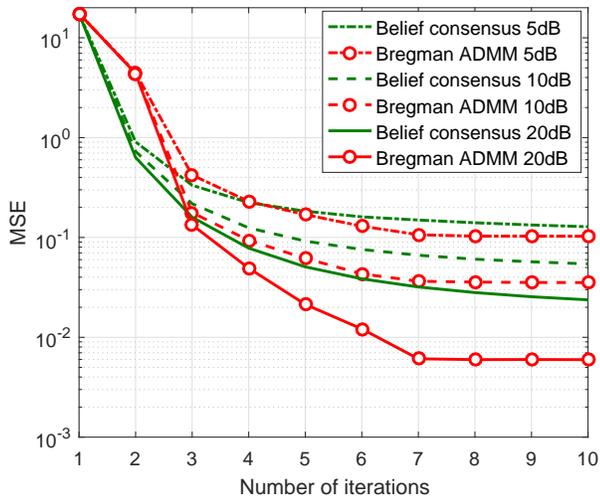}
\caption{MSE of parameters versus the number of consensus iterations.}\label{fig6}
\centering
\end{figure}

\edit{We further evaluate the performance of the proposed distributed cooperative detection algorithms in the condition of noisy inter-user links. \wj{In Fig.~\ref{fig7}, the BER performance of the proposed distributed algorithms versus $E_b/N_0$ is plotted, where the SNR corresponding to the inter-user links are set to be 10dB.
\editw{The number of consensus iterations is $p=10$. It is seen that due to the noisy inter-user links, the performance of both distributed schemes at $p=10$ cannot attain that of the centralized scheme.} We can also observe the Bregman ADMM-based method outperforms the belief consensus-based algorithm. To further analyze the convergence properties of the two distributed schemes,
in} Fig. \ref{fig6}, the mean squared error (MSE) of local parameters $\bm{\theta}_k$ versus the number of consensus iterations is illustrated. The MSE is defined as ${\sum_{k=1}^K \|\bm{\theta}_k-\bar{\bm{\theta}}\|_2^2}$ where $\bar{\bm{\theta}}=\frac{\sum_k \bm{\theta}_k}{K}$. \editw{Three \wj{SNR} scenarios of the inter-user links} are considered, i.e., $\textrm{SNR}=\{5, 10, 20\}$dB. As expected, larger SNR leads to smaller uncertainty and the MSE performance is better. Due to the vanishing factor ${\alpha^p}=\frac{1}{p}$, belief consensus-based algorithm converges slower than the Bregman ADMM-based method. Note that the MSE performance gap between them becomes even larger at higher SNR. This is due to the fact that belief consensus algorithm uses the same vanishing factor at high SNR while Bregman ADMM-based method benefits from small noise variance. Therefore, Bregman ADMM is more efficient in noisy \editw{inter-user} link networks.}

\section{Conclusions}
\revise{In this paper, we proposed factor a graph-based low-complexity message passing receivers for MIMO-SCMA system over frequency selective channels. Since the direct factorization of the joint posterior distribution leads to huge complexity in message updating, we introduced auxiliary variables and constructed a stretched factor graph. EP was employed to approximate the messages of data symbols to Gaussian distribution and a hybrid BP-EP receiver was proposed.
Considering the poor convergence property of the standard BP on loopy factor graph, we proposed to employ appropriate counting numbers to convexify the Bethe free energy and derived convergence-guaranteed BP-EP receiver.
We further considered a cooperative network and proposed two distributed cooperative detection schemes, i.e., belief consensus-based algorithm and Bregman ADMM-based method. The proposed iterative receivers were evaluated by Monte Carlo simulations and compared with the other schemes. It was shown that the proposed Stretch-BP-EP receiver performed close to the MMSE-based receiver with much lower complexity. The proposed Conv-BP-EP receiver outperforms the Stretch-BP-EP by improving the convergence property. Compared with the orthogonal multiple access counterpart, MIMO-SCMA system with the proposed receivers was shown to be able to support 50\% more users over frequency selective fading channels, with negligible BER performance loss. In cooperative networks, it was verified that BER performance could be further improved by exploiting the diversity gain using the proposed two distributed cooperative detection schemes. Moreover, compared with the belief consensus-based algorithm, Bregman ADMM-based method was shown to be more attractive in practical noisy inter-user links.}

\appendix [Derivation of Messages \eqref{ftoxcov} and \eqref{xtofcov}]

\revise{Solving the optimization problem \eqref{optim1} yields the corresponding beliefs as
\begin{align}\label{baaaa}
b_a(\tb{x}_a)&\propto f_a(\tb{x}_a)^{\frac{1}{c_a}}  \prod_{i\in\mathcal{N}(a)} \exp(\frac{\beta_{ai}(x_i)}{c_a})\\\label{biiii}
b_i(x_i)&\propto \prod_{f_a\in\mathcal{N}(i)} \exp(-\frac{\beta_{ai}(x_i)}{c_i}).
\end{align}
For clarity, we make the following definitions: $\tau_i=(1-c_i)/|\mathcal{N}(i)|$, $\mu_{x_i\to f_a} (x_i)=\exp (\frac{\beta_{ai}(x_i)}{c_a})$ and $\mu_{f_a\to x_i} (x_i)=b^{\tau_i}_i(x_i)\exp ({-\beta_{ai}(x_i)})$. Then we have
\begin{align}\label{message1}
\exp ({\beta_{ai}(x_i)})&=\mu^{c_a}_{x_i\to f_a} (x_i)\\\label{message2}
\exp ({-\beta_{ai}(x_i)})&=\mu_{f_a\to x_i} (x_i)\cdot b^{-\tau_i}_i(x_i).
\end{align}
Substituting \eqref{message1} and \eqref{message2} into \eqref{baaaa} and \eqref{biiii} yields,
\begin{align}\label{baaa1}
&b_a(\tb{x}_a)\propto f_a(\tb{x}_a) \prod_{i\in\mathcal{N}(a)} \mu_{x_i\to f_a} (x_i)\\\label{biii1}
&b^{c_i+\tau_i |\mathcal{N}(i)|}_i(x_i)\propto \prod_{f_a\in\mathcal{N}(i)} \mu_{f_a\to x_i} (x_i).
\end{align}
Then we define two auxiliary messages as
\begin{align}\label{message3}
\tilde{\mu}_{f_a\to x_i} (x_i)&\propto\int  f_a^ \frac{1}{c_a}(\tb{x}_a) \prod _{i^{'}\in \mathcal{N}(a)\backslash i}\mu_{x_{i^{'}}\to f_a} (x_{i^{'}}) \textrm{d} {x}_{i^{'}}\\\label{message4}
\tilde{\mu}_{x_i\to f_{a}} (x_i)&\propto \prod_{f_{a^{'}}\in \mathcal{N}(i)\backslash f_a} \mu_{f_{a^{'}} \to i} (x_i).
\end{align}
From \eqref{baaa1}-\eqref{message4}, we have
\begin{align}
b_i(x_i)&=\tilde{\mu}_{x_i\to f_{a}} (x_i) \mu_{f_a\to x_i} (x_i)\nonumber\\
&=\int b_a(\tb{x}_a) \textrm{d} \tb{x}_a\backslash x_i\nonumber\\
&= \tilde{\mu}_{f_a\to x_i} (x_i){\mu}_{x_i\to f_{a}} (x_i).
\end{align}
Comparing to ${\mu}^{c_a}_{x_i\to f_{a}} (x_i) \mu_{f_a\to x_i} (x_i)= b^{\tau_i}_i(x_i)$, we have
\begin{align}\label{message5}
{\mu}^{c_a}_{x_i\to f_{a}} (x_i) \mu_{f_a\to x_i} (x_i)=\tilde{\mu}_{x_i\to f_{a}}^{\tau_i} (x_i) \mu^{\tau_i}_{f_a\to x_i} (x_i)
\end{align}
and then
\begin{align}\label{message6}
{\mu}_{x_i\to f_{a}} (x_i)=\tilde{\mu}_{x_i\to f_{a}}^{\frac{\tau_i}{c_a}} (x_i) \mu^{\frac{\tau_i-1}{c_a}}_{f_a\to x_i} (x_i).
\end{align}
Based on \eqref{message5} and \eqref{message6}, we have
\\
\begin{align}\label{message7}
\mu_{f_a\to x_i} (x_i)=\tilde{\mu}_{x_i\to f_{a}}^{\frac{\tau_i-c_a}{c_a-\tau_i+1}} (x_i) \tilde{\mu}_{f_a\to x_i}^{\frac{c_a}{c_a-\tau_i+1}} (x_i).
\end{align}
Finally, we substitute \eqref{message7} into \eqref{message6} and obtain
\begin{align}\label{message8}
{\mu}_{x_i\to f_{a}} (x_i)=\tilde{\mu}_{x_i\to f_{a}}^{\frac{1}{c_a-\tau_i+1}} (x_i) \tilde{\mu}_{f_a\to x_i}^{\frac{\tau_i-1}{c_a-\tau_i+1}} (x_i).
\end{align}
With the definition of $\tau_i$, it is easy to see messages \eqref{message7} and \eqref{message8} are the same as \eqref{ftoxcov} and \eqref{xtofcov}.
}
\bibliographystyle{IEEEtran}
\bibliography{SCMA}
\end{document}